\documentclass{kapproc} 

\usepackage{epsfig}

\kluwerbib

\begin{document}

\articletitle{Multifrequency Strategies for the Identification of Gamma-Ray 
Sources}

\articlesubtitle{ }

\author{Reshmi Mukherjee}
\affil{Barnard College, Columbia University\\
Department of Physics \& Astronomy, New York, NY 10027}
\email{muk@astro.columbia.edu}

\author{Jules Halpern}
\affil{Columbia University\\
Department of Astronomy, New York, NY 10027}
\email{jules@astro.columbia.edu}

\begin{abstract}

More than half the sources in the Third EGRET (3EG) catalog have no firmly
established counterparts at other wavelengths and are unidentified. 
Some of these unidentified sources have remained a mystery since the first
surveys of the $\gamma$-ray sky with the COS-B satellite. The 
unidentified sources generally have large error circles, and finding
counterparts has often been a challenging job. A multiwavelength approach, 
using X-ray, optical, and radio data, is often needed to understand the nature
of these sources. This chapter reviews the technique of identification of EGRET
sources using 
multiwavelength studies of the gamma-ray fields. 
\end{abstract}


\section{Introduction and historical overview}

\begin{figure}[b!]
\centerline{\psfig{file=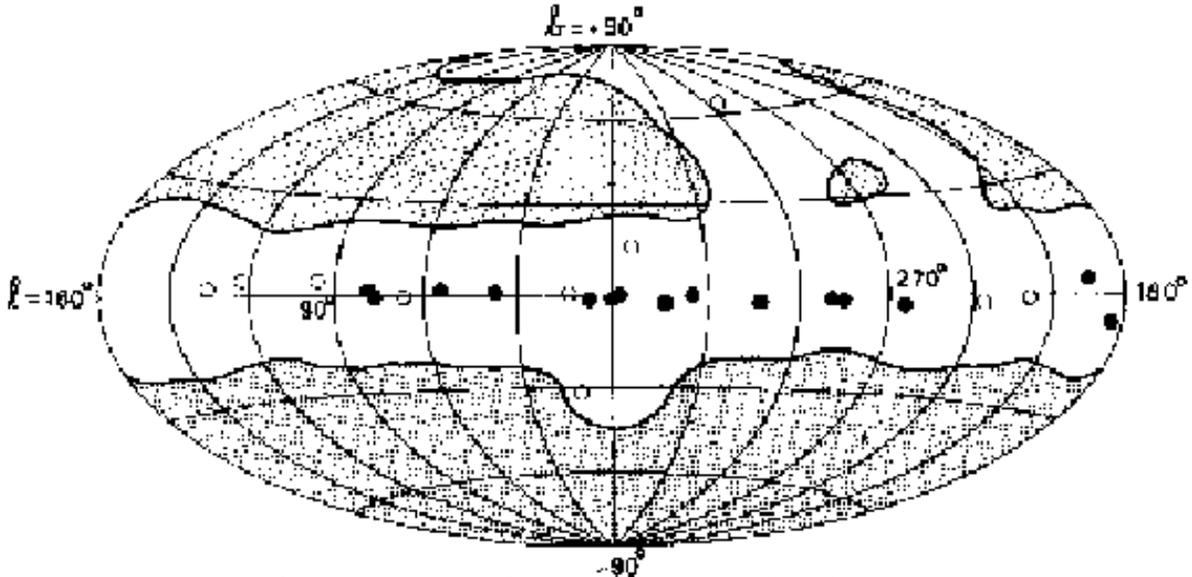,height=2.5in,bbllx=50pt,bblly=280pt,bburx=560pt,bbury=510pt,clip=.}}
\caption{ Point sources of $\gamma$-rays in the second and final COS-B 
catalog. Sources with flux brighter than 
$1.3\times 10^{-6}$ ph cm$^{-2}$ s$^{-1}$ are denoted with filled circles. 
Only the unshaded area represents the sky portion surveyed for point 
sources. Figure from 
(Swanenburg et al. 1981). } 
\label{unid}
\end{figure}

The discovery of point-like high energy sources in the $\gamma$-ray sky has been
one of the most exciting results in the field of $\gamma$-ray astronomy, since 
the advent of the first satellites in the 1970s. These sources have included 
exotic and energetic objects such as active galaxies, powered by super massive
black holes, pulsars, and powerful and mysterious $\gamma$-ray bursts, and have 
enabled  
us to explore some of the highest energy accelerators in the cosmos. But, 
perhaps the most mysterious and enigmatic of the sources have been the 
``unidentified'' $\gamma$-ray sources. As the qualifier suggests, these are 
objects in the $\gamma$-ray sky with no identifications or known 
counterparts at other wavebands. Some of the unidentified sources have 
remained so since the first surveys of the $\gamma$-ray 
sky carried out by the COS-B satellite in the 1970s. As described in Chapter 1
of this book, COS-B detected a total of
25 sources, of which only the pulsars, Crab and Vela, the molecular cloud 
$\rho$-Oph and the first extragalactic source, 3C 273 were identified (Bignami
\& Hermsen 1983). The
remaining 21 sources in the 2nd COS-B catalog had no unambiguous counterparts 
at other wavebands. Figure~\ref{unid} shows the COS-B skymap. As one of the first
catalogs of $\gamma$-ray sources it represents a 
significant step in the field of $\gamma$-ray astronomy. 

\begin{figure}[t!]
\centerline{\psfig{file=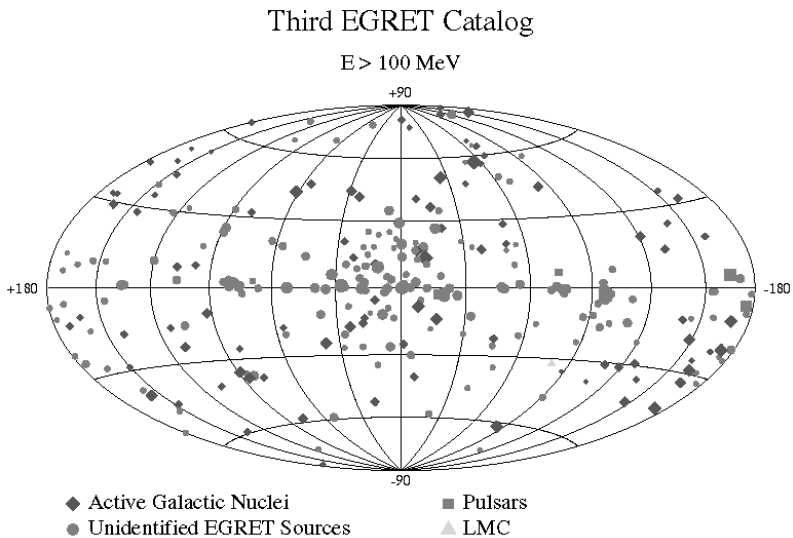,height=2.8in,bbllx=-80pt,bblly=135pt,bburx=694pt,bbury=658pt,clip=.}}
\caption{Point sources (from the Third EGRET Catalog) detected by EGRET at 
$> 100$ MeV. The size of the symbols are scaled 
according to source flux. The unidentified sources are shown as filled 
circles. Figure from (Hartman et al. 1999).}
\label{egretskymap}
\end{figure}

Following COS-B, the next major step in $\gamma$-ray astronomy came with the 
launch of the {\sl Compton Gamma Ray Observatory} (CGRO) 
in 1991, when the on-board EGRET (Energetic Gamma-ray Experiment Telescope) 
instrument carried out improved surveys of the $\gamma$-ray sky, 
at relatively better angular resolution. EGRET's success was
tremendous, and a total of 271 point sources of high energy $\gamma$-rays above 
100 MeV, were catalogued (Hartman et al. 1999). 
However, only a fraction of these sources were 
identified. The unidentified sources comprised the majority of the $\gamma$-ray 
point sources, some in fact being originally discovered by the COS-B 
satellite. The nature of these persistent 
$\gamma$-ray sources is an outstanding mystery in high energy astrophysics, in
some cases almost three decades after their discovery. 
Figure~\ref{egretskymap} shows the point sources catalogued in the Third EGRET (3EG) catalog,
detected above 100 MeV. The unidentified sources, shown as filled circles,
constitute the largest class of the EGRET sources.  Resolving the mystery of 
the $\gamma$-ray sources is a significant challenge across all wavebands in
astronomy. A nice recent review of the current status in the quest for the
identification of the high energy $\gamma$-ray sources is given by 
Caraveo (2002). 

\subsection{EGRET source sensitivity}

It is important to point out that EGRET did not survey all regions of the sky
with the same sensitivity. Figure~\ref{egretexpo} shows the sky exposure for EGRET above 100
MeV for the duration of the EGRET mission. 
The significance of detection $S$ of a
source with flux $F$ is related to the exposure $E$ and background $B$ by 
$S\sim F\sqrt{E/B}$. Since the EGRET intensity map is dominated
by strong diffuse emission along the Galactic plane (Hunter et al. 1997),
the $\gamma$-ray source detection threshold is definitely higher in regions of 
low exposure or high diffuse radiation. 
Because of the larger systematic uncertainties in the EGRET analysis for the
high intensity Galactic plane region, the 3EG catalog actually adopts two
different and separate criteria for source detection thresholds. A source is
listed in the catalog if it is detected at $4\sigma$ or higher for
$|b|>10^\circ$, and $5\sigma$ or higher for $|b|<10^\circ$. Because of the
differences in source sensitivities, the EGRET catalog cannot be taken as a
uniform sampling of the $\gamma$-ray sky, and this has to be taken into account
in all source population studies.  

\begin{figure}[b!]
\begin{center}
\vskip 2.8in
{\bf Figure and complete paper with good quality graphics available at: 
http://www.astro.columbia.edu/$\sim$muk/mukherjee\_multiwave.pdf}
\caption{ EGRET sky exposure in units of $10^8$ cm$^2$s for photon energies $>
100$ for the sum of {\sl CGRO} 1, 2, 3 and 4 (1991 April - 1995 October). The
intervals of contour spacing are $2\times 10^8$. }
\end{center}
\label{egretexpo}
\end{figure}

\subsection{Source distributions of the unidentified sources}

EGRET measured the source location, the $\gamma$-ray light curve and the 
spectra of the individual $\gamma$-ray sources. Typical EGRET observations lasted
for a period of about 2 weeks, although some observations were as short as a 
week, while others were as long as 3 to 5 weeks. 
EGRET's threshold sensitivity ($> 100$ MeV) for a 
single 2-week observation was 
$\sim 3\times 10^{-7}$ photons cm$^{-2}$ s$^{-1}$. Details of the EGRET
instrument, and data analysis techniques are given elsewhere 
(Thompson et al. 1993; Hartman et al. 1999). 

Source distributions of the unidentified EGRET sources are often useful in
understanding the overall properties of these sources, particularly in 
providing a constraint on the average distances or luminosities of the 
sources. One of the first studies of the unidentified source distributions 
as a
function of Galactic latitude and longitude was carried out by Mukherjee et
al. (1995), using the source lists available at that time. The unidentified
Galactic sources were found to have an average distance between 1.2 and 6 kpc,
and isotropic luminosities in the range $(0.7-16.7) \times 10^{35}$ erg
s$^{-1}$. These results were in agreement with the earlier findings of Bignami
\& Hermsen (1983) for the COS-B data. 

\begin{figure}[b!]
\centerline{\psfig{file=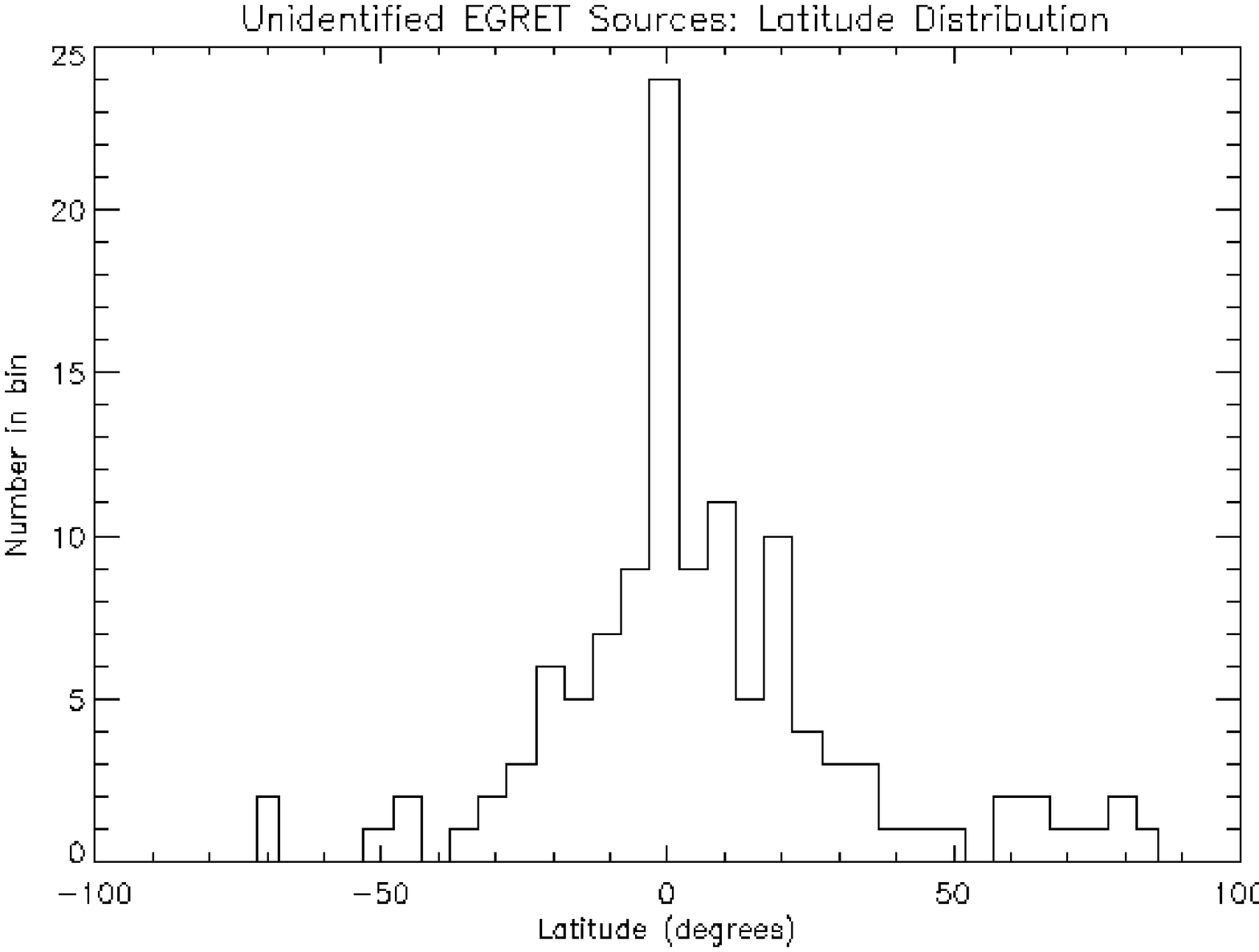,height=2.3in,bbllx=-2pt,bblly=177pt,bburx=615pt,bbury=615pt,clip=.}}
\centerline{\psfig{file=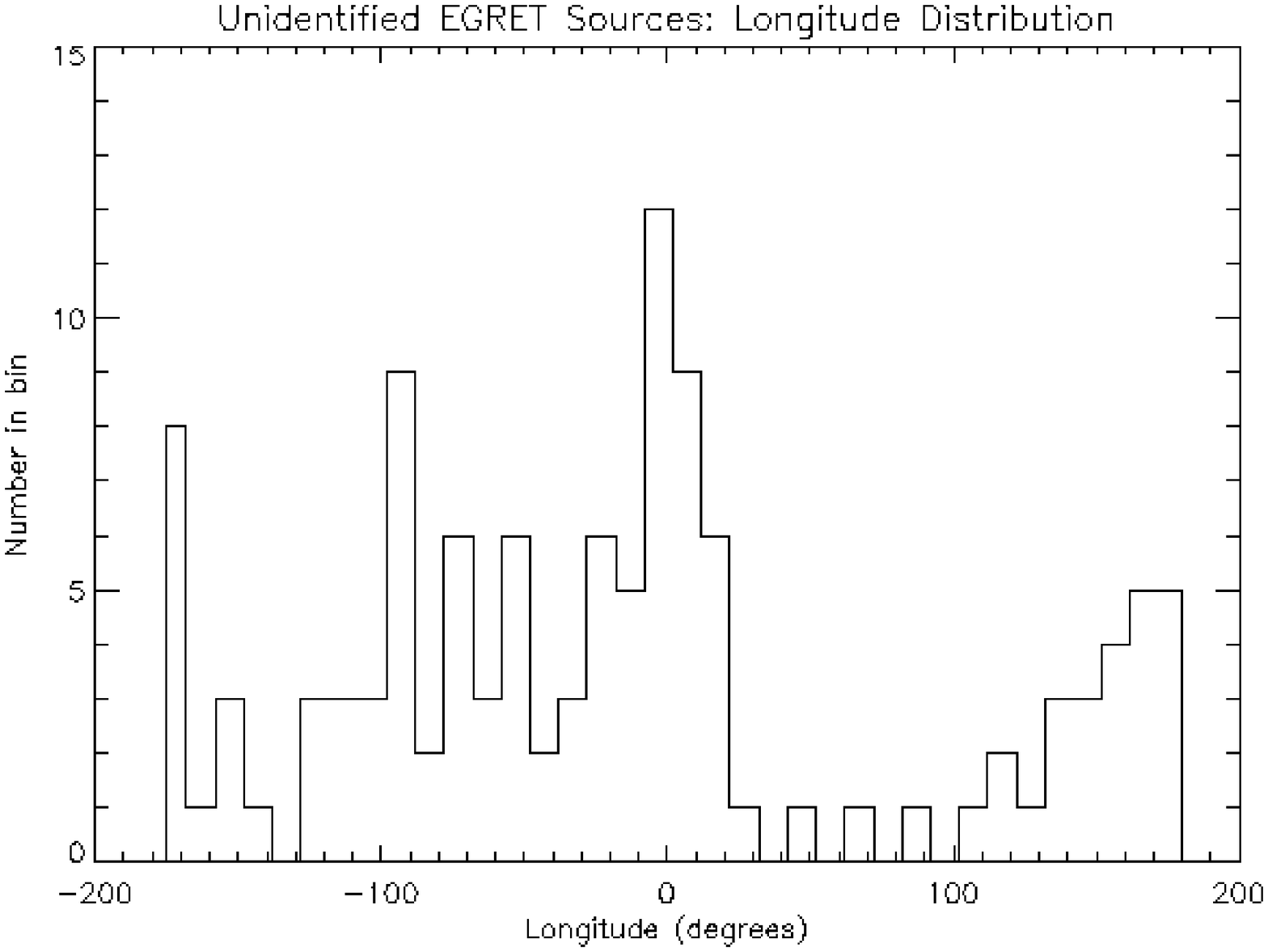,height=2.3in,bbllx=-2pt,bblly=177pt,bburx=615pt,bbury=615pt,clip=.}}
\caption{Distributions in (top) latitude and (bottom) longitude of all the
unidentified sources in the 3EG catalog.}
\label{latlong}
\end{figure}

Figure~\ref{latlong}
shows the latitude distribution of all the unidentified sources in 
the 3EG catalog. In terms of source counts, 90\% of the EGRET sources at 
$b<10^\circ$ are unidentified. At high latitudes, $b>10^\circ$ where a large 
number of the EGRET sources are identified as blazars, the fraction of the 
unidentified sources is 50\%. Gehrels (2000) and Grenier (2000) note that 
there is an excess of faint sources at mid-latitudes, $10^\circ<b<30^\circ$ 
that are fainter and softer than the low latitude sources, on the average. 
It has been suggested that these mid latitude sources could possibly be
associated with the Gould Belt structure (Gehrels 2000; Grenier 2000). 
Figure~\ref{latlong} (bottom) 
shows the longitude distribution of the unidentified sources in 
the 3EG catalog. 

Log $N$-log $S$ studies of EGRET sources are often useful in learning about the
general characteristics of EGRET source populations. One of the first such
studies was carried out by Oz\"el \& Thompson (1996) for comparing unidentified
EGRET sources and EGRET-detected AGN populations. Similarly, Reimer \& Thompson
(2001) have studied log $N$-log $S$ distributions for 3EG sources. 
Population studies of EGRET sources, taking advantage of source distributions
and correlations have been used to infer the nature of EGRET unidentified
sources. We have not summarized these studies in this article, but we point to
several review articles that describe these in some detail (Mukherjee, Grenier
\& Thompson 1997; Caraveo 2002). 

\subsection{Counterpart searches - challenges in the identification process}

EGRET's better sensitivity and superior angular resolution in comparison to
COS-B led to nearly a ten-fold increase in the number of $\gamma$-ray source
detections over COS-B. However, this did not necessarily lead to an increase 
in the number of source identifications. 
The identification of the EGRET sources, particularly those close to the
Galactic plane has proved to be challenging. The error box of the typical 
EGRET source is large, $\sim 0.5^\circ-1^\circ$, and identifications and
counterpart searches on the basis
of position alone has been difficult. This is further hampered for the low
latitude sources by the 
presence of bright Galactic diffuse emission along the plane. Also, a lack 
of tight correlation between the $\gamma$-ray flux and other properties,
like X-ray flux, core radio flux, etc., allows only the strongest sources to 
be identified on the basis of position alone.  

Counterpart searches of $\gamma$-ray sources usually start with looking 
for
``more of the same'' kinds of sources.  
So far the identified sources fall into two major source classes: 
blazars and pulsars. Most of the 
blazar identifications are at high Galactic latitudes, where the source fields
are less crowded, positions are better determined, and additional resources
such as $\gamma$-ray flux variability, correlated with variability at radio or
optical bands make the identifications more confident. 
All the pulsars detected by EGRET are at low latitudes. It is therefore quite
likely that at least a fraction of the unidentified sources at $b<10^\circ$ 
will belong to the pulsar class. In this case, a definite time signature will 
be needed in the $\gamma$-ray data, which was usually difficult in the case of
EGRET data. 
Similarly, it is likely that a large fraction of the high latitude 
unidentified sources, with better source positions obtained in the future with
GLAST, will turn out to be associated with blazars (see the chapter by Torres).

An ``elusive template'' for possibly
another class of $\gamma$-ray source is provided by Geminga, the only
radio-quiet pulsar in the EGRET data (see Caraveo, Bignami \& Tr\"umper
1996 for a review). 
Although Geminga is probably the nearest member of this class 
(see also discussion on 3EG J1835+59 below), it is possible that other
candidates will be found in the era of GLAST. In fact, some of the fainter, 
mid-latitude EGRET sources (more local Galactic population) could be accounted
for by Geminga-like pulsars (Gehrels et al. 2000). 

For a $\gamma$-ray source that does not definitely 
belong to the blazar or pulsar category, 
a search for counterpart usually relies on one of two techniques. Generally, 
identifications of $\gamma$-ray sources are carried out using either the help 
of 
population studies or on a case-by-case basis relying on information based on
multiwavelength observations. In the former, $\gamma$-ray source distributions 
and properties of populations of $\gamma$-ray sources are compared with 
properties of other source classes. 
In the latter case, error boxes of individual $\gamma$-ray
sources are studied using information obtained at other wavebands. This 
chapter attempts to summarize the multiwavelength approach to the 
identification of $\gamma$-ray sources. 

\subsection{The multiwavelength approach}

Studying the optical to X-ray data of 3EG unidentified sources has in several
cases shed some light on the nature of the EGRET source. This approach has now 
been applied successfully to several of the EGRET sources. The first steps in
this process usually involve the study of archival {\sl ASCA} or {\sl ROSAT}
data of the EGRET fields, with a follow up of optical and/or radio 
observations of X-ray sources in the error boxes. One of the first 
exhaustive studies of this kind was carried out by Roberts, Romani \& Kawai (2001) who
presented a catalog of {\sl ASCA} images in the 2-10 keV band of fields
containing bright EGRET sources. Although time consuming, this
``case-by-case'' method has met with success in several cases. In the following
we describe some of the individual cases, discussed in no particular order. 

\section{Blazars and EGRET unidentified sources}

The majority of the identified EGRET sources are blazars (flat-spectrum radio
quasars and BL Lac objects) - the only kinds of AGN 
that EGRET has detected with any measure of confidence. 
Mattox et al. (1997)
and Mattox, Hartman, \& Reimer (2001) have studied the statistical issues
concerning the identification of EGRET sources with blazars, and have presented
the probabilities of association of individual sources with blazars. In the 3EG
catalog Mattox et al. (2001) find that 46 EGRET sources may be confidently
identified with blazars, while an additional 37 are plausibly identified with
radio sources. 

The blazars seen by EGRET all share several common characteristics: they are 
radio-loud, flat spectrum sources, with radio spectral indices 
$0.6>\alpha>-0.6$ (von Montigny et al. 1995). Most of the EGRET sources confidently identified with blazars are
charcterized by strong radio fluxes ($> 500$ mJy) at 5 GHz. 
EGRET blazars have a continuum 
spectrum that is non-thermal, and are characterized by 
strong variability and optical polarization. 
In counterpart searches of unidentified EGRET sources, the EGRET source is
usually examined to see if it fits the blazar template. Here we describe
multiwavelength studies of EGRET fields that have led to the identification of
the EGRET source with a blazar. 

\subsection{ A Blazar counterpart for 3EG J2016+3657}

This is an example of a low-latitude EGRET source, 3EG J2016+3657, that was
identified with a blazar behind the Galactic plane, B2013+370. 
Although rare, it is certainly not unexpected that several
of the ``Galactic'' unidentified sources will turn out to be blazars, given the
isotropic distribution the of $\gamma$-ray blazar population.  

3EG J2016+3657 was identified after a detailed 
study was carried out of 
archival X-ray data, with follow-up optical observations of the 
the $\gamma$-ray error box (Mukherjee et al. 2000). The identification was 
soon confirmed by Halpern et al. (2001a) who concluded that B2013+370 was the 
most likely counterpart, after optical spectroscopic identifications of all 
soft and hard X-ray sources in the error circle of the EGRET source eliminated
the other candidates. We discuss these results here in some detail in order to
illustrate the multiwavelength ``strategy'' of the identification of 3EG
sources. 

\begin{figure}[t]
\vspace{-0.5cm}
\begin{minipage}{0.96\linewidth}
  \begin{center}
    \includegraphics[height=18pc]{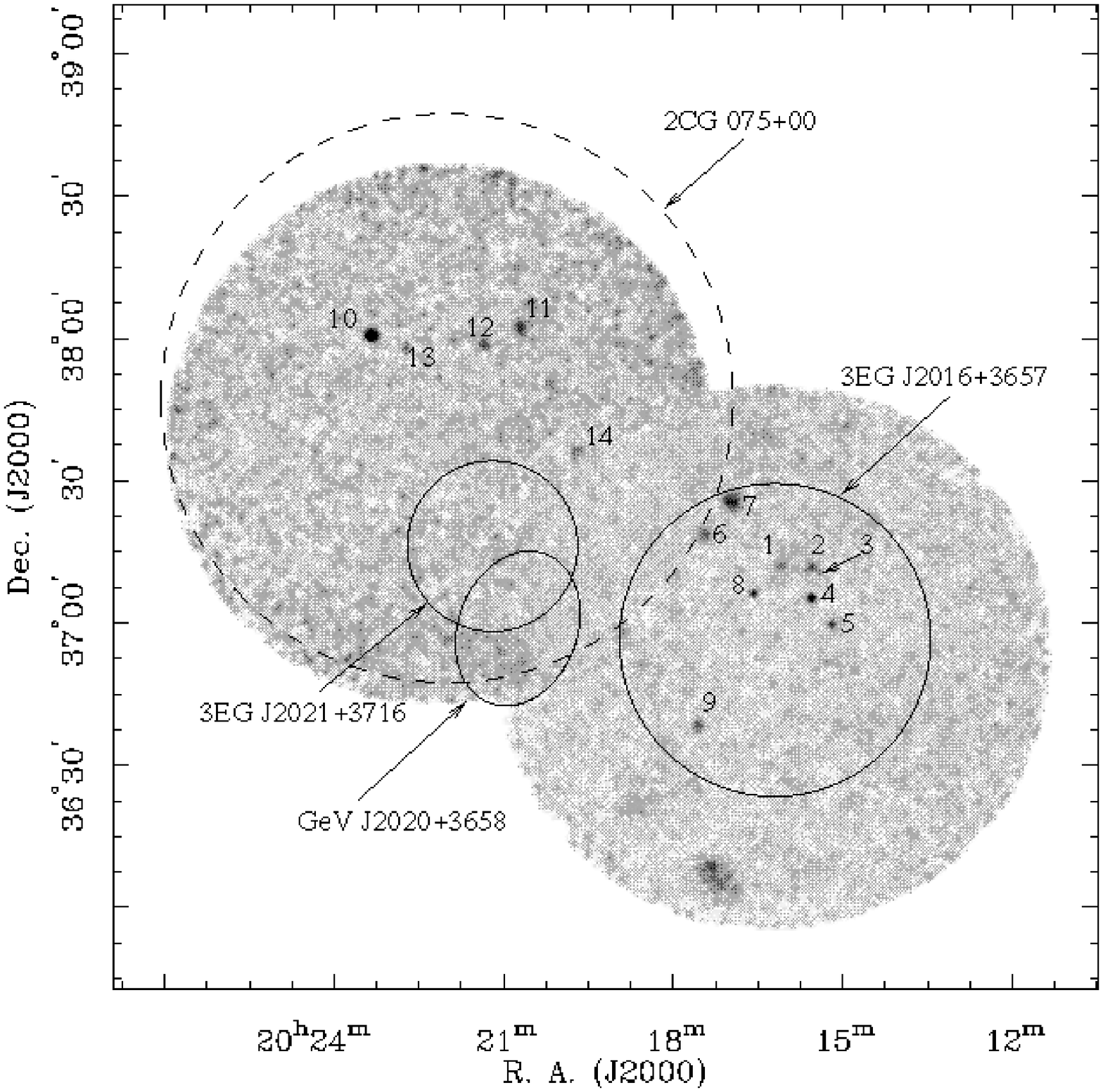}
    \includegraphics[height=18pc]{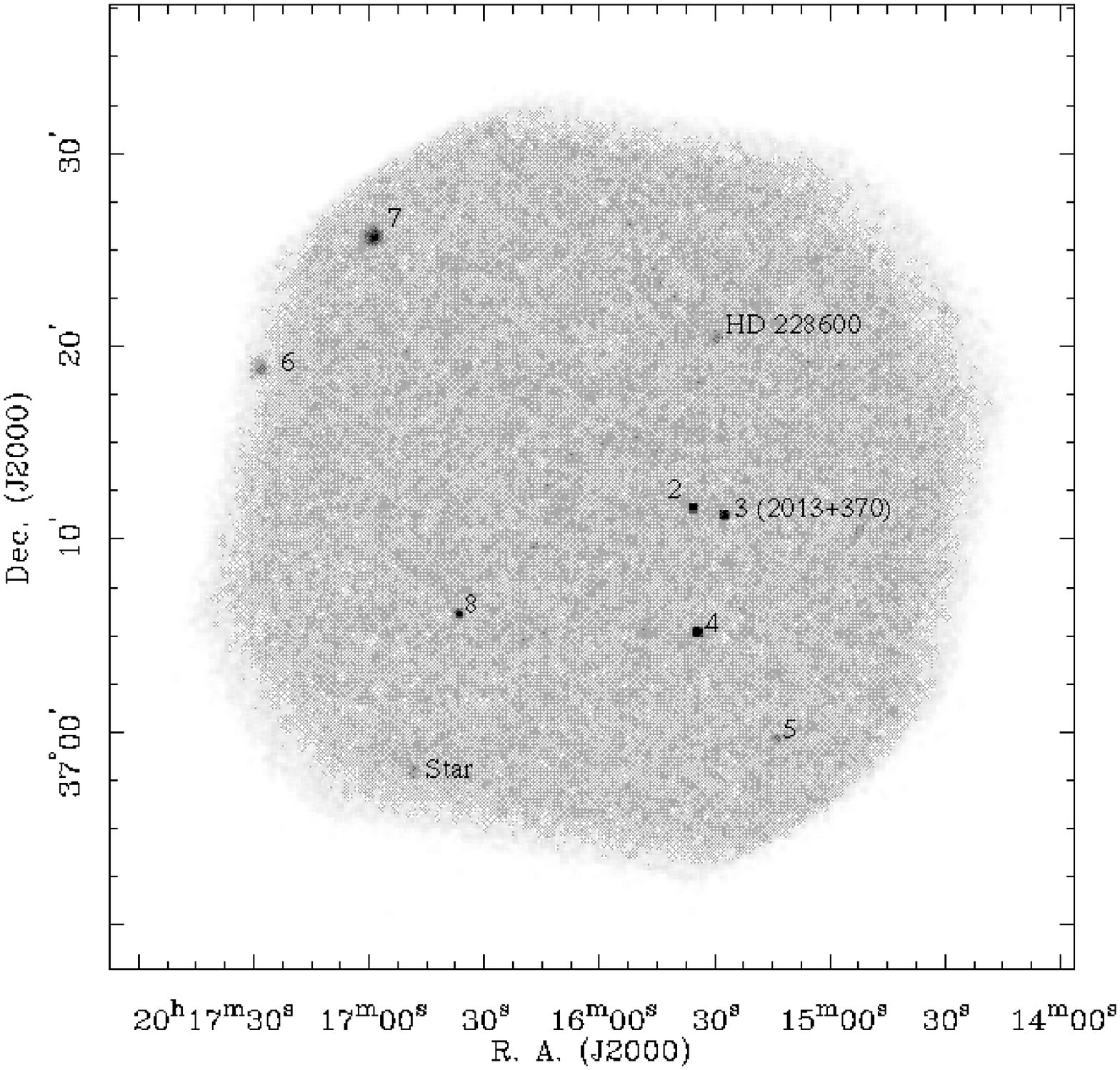}
  \end{center}
\vspace{-0.5cm}
  \caption{(Top) $ROSAT$ soft X-ray image of 3EG J2016+3657 and 3EG J2021+3716. 
The circles for the 
two 3EG sources correspond to the $\sim 95$ \% confidence contours. The dashed 
circle corresponds to the COS--B source 2CG 075+00. The GeV Catalog source 
(Lamb \& Macomb 1997) is also shown. The minimum detectable
intrinsic flux for the $ROSAT$ image was $6.5\times 10^{-13}$ erg cm$^{-2}$
s$^{-1}$. (Bottom) ROSAT HRI X-ray image of the field around 3EG J2016+3657. The image
shows the sources 2 and 3 (B2013+370) as clearly resolved point sources. Both
figures are from Mukherjee et al. (2000).}
\label{2016rosatpspc}
\end{minipage}
\end{figure}

3EG J2016+3657 \& 3EG J2021+3719 are two sources in the Cygnus region 
probably associated with the unidentified COS--B source 2CG 075+00 
(Pollack et al. 1985). 
The error circles of both 3EG J2016+3657 \& 3EG J2021+3716 are covered by 
archival X-ray imaging observations with {\it ROSAT} (PSPC and HRI) and 
{\it ASCA}, as well as {\sl Einstein} IPC (Wilson 1980). Figure~\ref{2016rosatpspc} shows the {\it ROSAT} 
soft X-ray ($0.2 -2.0$ keV) and HRI image of the region, along with the EGRET
error circles. The X-ray point source 
positions, marked in the figure, derived from the {\it ROSAT} analysis were used to search
for counterparts to the X-ray sources. 

Halpern et al. (2001a) used the MDM 2.4~m and the KPNO 2.1~m
telescopes to obtain a complete set of optical identifications of all 
X-ray point sources within the error circles of the two EGRET sources. It turns
out that other than source \# 1 and \# 3 in figure~\ref{2016rosatpspc}, the 
other sources in the EGRET fields are either cataclysmic variables (CVs), or
Wolf-Rayet stars or binary O stars, all unlikely to be $\gamma$-ray emitters. 
(Note, however, under some circumstances Wolf-Rayet binaries are expected to 
be significant gamma-ray emitters (e.g. Benaglia \& Romero 2003). Possible high
energy emission of early type binaries is also discussed in the chaper by Rauw
in this book). 
The two sources of interest in the field are the
supernova remnant (SNR) CTB 87 (source \# 1) and the blazar-like radio source
B2013+370 (source \#3). Of the two the blazar B2013+370 was suggested as the
most likely candidate. The other source, CTB 87, is too weak and too far away 
to be the likely candidate, and was therefore disfavored (see Halpern et
al. 2001a; Mukherjee et al. 2000 for details. However, a revised distance to 
CTB 87 places it half as far as previously believed (Kothes et al. 2003), which
weakens this argument slightly.) 

Other characteristics of B2013+370 supports the identification with 3EG
J2016+3657. B2013+370 has all the blazar-like characteristics of typical 
EGRET identifications - compact, 
extragalactic, non-thermal radio source, variable at optical and mm (90 GHz, 
142 GHz) wavelengths, with a 5 GHz flux of $\sim 2$ Jy. The spectral energy
distribution (SED) of 3EG J2016+3657 is characterized by a synchrotron peak at
lower energies, a Compton peak at higher energies, with most of the power 
output in $\gamma$-rays and confirms the blazar nature of the source. All 
these observations
suggest that 3EG J2016+3657 fits the blazar template, and that 
B2013+370 is the identification for the EGRET source.

\subsection{ 3EG J2027+3429: Another blazar behind the Galactic plane?} 

\begin{figure}[t]
\vspace{-0.5cm}
\begin{minipage}{0.96\linewidth}
  \begin{center}
     \includegraphics[height=18pc]{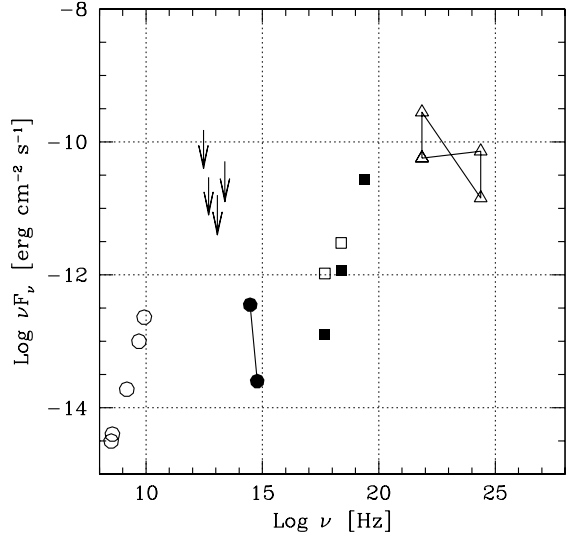}
  \end{center}
\vspace{-0.5cm}
  \caption{Spectral energy distribution (SED) of 3EG J2027+3429, assuming that it
is associated with the X-ray source WGA J2025.1+3342. The symbols are as
follows: open circles - radio, filled circles - optical, open and filled 
squares - BeppoSAX, triangles - EGRET. The arrows correspond to IRAS
upper limits. Note the synchrotron and inverse Compton humps characteristic of
EGRET blazars. Figure from Sguera et al. (2003).   }
\label{sguera2027}
\end{minipage}
\end{figure}

3EG J2027+3429, also in the Cygnus region, has been recently suggested to be
another blazar behind the Galactic plane. Using a multiwavelength strategy,
Sguera et al. (2003) have suggested the 
BeppoSAX X-ray source WGA J2025.1+3342, 
to be associated with the EGRET source. A search for X-ray counterparts in the
EGRET error box using archival BeppoSAX data yielded several X-ray point
sources, with WGA J2025.1+3342 being the strongest. WGA J2025.1+3342 
is also highly variable
at X-ray energies, and has a flat spectrum in the range 1-100 keV. 
A cross-correlation of 
these X-ray sources with radio catalogues found
only two of the X-ray point sources in the EGRET error circle to be 
associated with radio sources, with WGA J2025.1+3342 being the brightest 
radio source. At radio wavelengths, the source was found to have a flat 
spectrum in the range 0.3-10 GHz, and is a bright, compact object. 
Optical observations of the source by Sowards-Emmerd et al. 
(2003) suggest that the spectrum has emission lines of the Balmer series, and is
therefore a quasar at $z=0.219$. All these
characteristics point towards a blazar identification of 3EG
J2027+3429. Figure~\ref{sguera2027} shows the SED of 3EG J2027+3429, assuming
the identification is the correct one. The SED is typical of a low-frequency
peaked blazar, with the synchrotron peak at mm/far IR range and the inverse 
Compton peak at $\gamma$-ray energies (Sguera et al. 2003). 
Once again, the analysis of archival radio, IR, optical and new X-ray
observations has suggested an identification for an EGRET unidentified
source. If correct, this is the second $\gamma$-ray blazar behind the Galactic
plane, and is very likely not to be the last. 

\subsection{3EG J2006-2321: A blazar with a weak radio flux} 

\begin{figure}[t]
\vspace{-0.5cm}
\begin{minipage}{0.96\linewidth}
  \begin{center}
     \includegraphics[height=8cm]{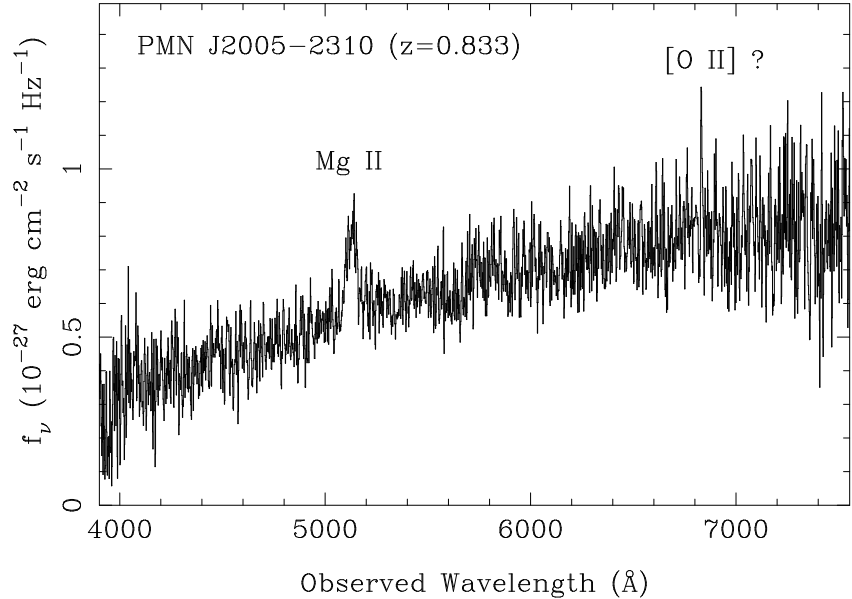}
  \end{center}
\vspace{-0.5cm}
  \caption{Spectrum of the optical counterpart of PMN J2005-2310, suggested as 
the counterpart of 3EG J2006-2321, from the KPNO 2.1 m. Analysis of archived
radio, X-ray data together with optical spectroscopy and polarimetry of the
region helped in the identification of the 3EG source. Figure from Wallace et al. (2002).   }
\label{wallace2006}
\end{minipage}
\end{figure}

Yet another blazar identification was made for the EGRET source 3EG J2006-2321
(Wallace et al. 2002), using a similar multiwavelength approach. The source was
identified with the flat-spectrum radio quasar PMN J2005-2310, after a careful
study of the field at radio, optical and X-ray energies. Its optical
counterpart has $V=19.3$ and $z=0.833$. 
Figure~\ref{wallace2006} shows the spectrum of PMN J2006-2310 from KPNO
2.1 m. Interestingly, this source has a 5 GHz flux density of 260 mJy, which is
the lowest of the 68 identified blazars in the 3EG catalog. Although this is
atypical of most EGRET blazar identifications (bright, $\sim 1$ Jy, radio
sources at 5 GHz), the identification is still plausible because the radio to
$\gamma$-ray flux density ratio is comparable to the ``confident'' blazar
identifications (see Figure~\ref{julesradio} and the discussion  in \S 2.4). As
Wallace et al. rightly point out, other weaker EGRET unidentified sources are
likely to be identified with low flux density radio sources in the future. 

\subsection{Blazars in the northern sky}

The $\gamma$-ray blazar content of the northern sky was recently explored by
Sowards-Emmerd et al. (2003), who used radio survey data to re-evaluate
correlations of flat spectrum radio sources with the EGRET
sources. This is similar to the approach traditionally used for the 
selection of blazar candidates in the past for EGRET sources (Hartman et
al. 1999 in the 3EG catalog; Mattox et al. 2001). Sowards-Emmerd et al. 
additionally carried out follow-up optical
spectroscopic observations with the Hobby Eberly Telescope (HET) to confirm 
the AGN candidate. This survey has resulted in the confirmation of the 
existing EGRET blazars and suggested blazar candidates for several 3EG
unidentified sources in the northern sky. If confirmed, the association of 3EG
sources at $b> 10^\circ$ with blazar-like radio sources is found to be 70\%. 
Unlike previous associations of EGRET sources with bright, 1 Jy radio sources
(Hartman et al. 1999), Sowards-Emmerd et al. have suggested 
plausible counterparts 
down to fluxes of $\sim 100$ mJy at 8.4 GHz. It is likely that in the future
GLAST era, better multiwavelength follow-ups will result in the association of
more 
$\gamma$-ray sources with weaker ($< 100$ mJy) radio sources. In that case, the
really interesting question will be what is the nature of the ``non-blazar'' EGRET
sources. 

\begin{figure}[t!]
\begin{center}
\includegraphics[height=8cm]{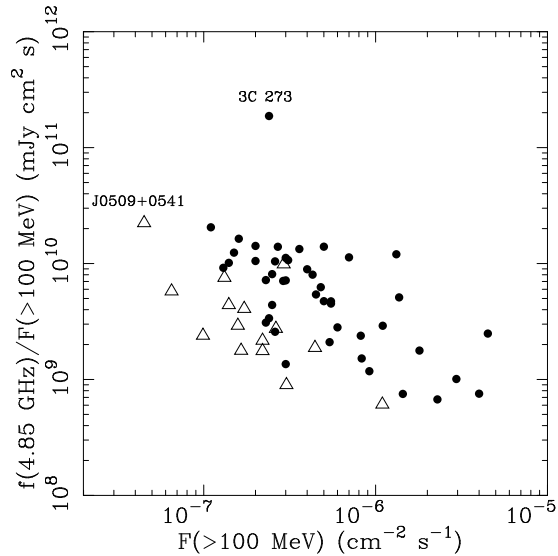}
\end{center}
\caption{Ratio of radio (4.85 GHz) flux density to the peak $\gamma$-ray flux of
the confident EGRET blazar identifications (circles) compared with that
for the 16 3EG sources tentatively identified with blazars (triangles). Note
that the marginally identified blazars are still plausible identifications,
although they have lower radio fluxes, as they fall within the same
radio/$\gamma$-ray flux ratio. Figure from Halpern et al. (2003).   }
\label{julesradio}
\end{figure}

Another multiwavelength study of ``lower confidence'' $\gamma$-ray blazars in the
3EG catalog was carried out by Halpern et al. (2003), who identified optical
counterparts of 16 3EG sources associated with blazars and obtained nine
redshifts. In each of these cases
very little optical information was previously available. 
Although
the radio identification of EGRET sources are not flux limited, because of
source confusion due to the large EGRET error circles, only the brightest radio
sources ($> 500$ mJy) are secure identifications. Figure ~\ref{julesradio} compares the radio
and $\gamma$-ray fluxes of the high confidence blazar identifications of Mattox et
al. (2001) with that of the 16 3EG sources studied by Halpern et al. (2003). 
These 16
blazars have lower radio fluxes than the
high-confidence blazar identifications, but 
are still plausible counterparts as they have the same radio to $\gamma$-ray flux
ratios. It is likely that many of the unidentified 3EG sources are blazars with
lower radio fluxes. In fact, this was the case for the AGN identification of 3EG
J2006-2321 discussed earlier. 

\subsection{Blazars in the southern sky}

On a smaller scale, Tornikoski et al. (2002) 
have carried out high frequency radio
observations at 90 and 230 GHz of a dozen 3EG sources in the southern 
hemisphere that were tentatively identified as blazars in the 3EG catalog. 
These radio observations have confirmed 5 of the sources as blazars. An
additional 4 unidentified EGRET sources have been identified as likely blazars,
based on their activity at mm wavelengths. 

\section{EGRET sources and radio galaxies}

Other than 
blazars, the only extragalactic sources to have been detected by EGRET are 
the radio galaxy Cen A, and the normal galaxy LMC. Radio galaxies are not known
to be strong $\gamma$-ray emitters. In the 3EG
catalog, Cen A (NGC 5128) is the only radio galaxy to be identified with an
EGRET source at energies above 100 MeV (Sreekumar et al. 1999), and provides
the first clear evidence that an AGN with a large-inclination jet can be
detected at $\gamma$-ray energies above 100 MeV. This is unlike the EGRET
blazars which are believed to have jets nearly aligned along our 
line-of-sight. Cen A's jet is offset by
an angle of about $70^\circ$ (Bailey et al. 1986; Fujisawa et al. 2000). 
Cen A is also a weak $\gamma$-ray source and has a 
derived $\gamma$-ray luminosity weaker by a factor of $10^{-5}$ compared to the
typical EGRET blazar.  Cen~A was probably detected by EGRET as it  
is the brightest and nearest 
radio galaxy ($z=0.0018$, $\sim 3.5$ Mpc). 
Cen A was the only one of its kind in the 3EG catalog, until recent reports of
a couple of other candidate radio galaxies to be identified with EGRET sources
(see below). It is very likely that the 
detection of more radio galaxies by EGRET has been 
limited by its threshold sensitivity. If this is true then 
there exists the exciting 
possibility that instruments like GLAST, with much higher
sensitivity, will detect more radio galaxies in the future.   

\subsection{3EG J1621+8203: The radio galaxy NGC~6251?}

In an effort to investigate the nature of the EGRET source 3EG J1621+8203, 
Mukherjee et al. (2002) have again used a multiwavelength approach 
and examined X-ray images of the
field from {\sl ROSAT} PSPC, {\sl ROSAT} HRI, and {\sl ASCA} GIS, as well as
radio and optical surveys 
with follow-up optical spectroscopic classification of
active objects within the error ellipse of the EGRET source. Except for one, 
all X-ray sources in the EGRET error box were identified with ordinary QSOs or
coronal emitting stars, all unlikely to be counterparts of the $\gamma$-ray 
source. The most notable object in the $\gamma$-ray error box is the bright FR I 
radio galaxy NGC 6251, which Mukherjee et al. (2002) have suggested as a
plausible counterpart for 3EG J1621+8203. 

As in the case of Cen A, 3EG J1621+8203 has a lower $\gamma$-ray luminosity 
($3\times 10^{43}$ ergs/s) than that of other
EGRET blazars (typically $10^{45}$ to $10^{48}$ ergs/s). 
Compared to Cen A, NGC 6251 is much further away ($z=0.0234$), which raises the
question whether it is luminous enough to have been detected by EGRET. 
However, NGC 6251
is most likely still detectable by EGRET because of its smaller jet angle
($45^\circ$) in comparison to that of Cen A ($70^\circ$). 
 
If 3EG J1621+8203 corresponds to NGC 
6251, then it would be the second radio galaxy to be detected in high energy
$\gamma$-rays. NGC 6251 is a notable candidate because of the possible link 
between FR I radio galaxies and BL Lac objects; FR I radio galaxies are
hypothesized to be the likely parent population of BL Lac objects (Urry \&
Padovani 1995). 

\subsection{3EG J1735-1500: Another new radio galaxy}

Yet another possible radio galaxy counterpart to an EGRET source was recently
suggested by Combi et al. (2003) for 3EG J1735-1500. 
In this case, the NRAO VLA Sky Survey (NVSS) (Condon et al. 1998) was used to
examine the radio sources within the 95\% EGRET error box. The radio galaxy
J1737-15 was suggested as the most likely counterpart. 
Combi et al. noted, however,  that another likely counterpart of 
3EG J1735-1500 could be the flat-spectrum, compact weak radio 
source PMN J1738-1502, also located in the error box. The lack of a unique counterpart
for an EGRET source, following a multiwavelength survey of the error box is not
surprising. In fact, this illustrates the problems associated in the 
counterpart 
searches for EGRET sources, which typically have large $\sim 1^\circ$ error
boxes. In this case, future observations with GLAST will help confirm the
identification for 3EG J1735-1500. 

\section{Radio quiet isolated neutron stars}

Caraveo (2002) has referred to isolated neutron stars (INS) as  ``elusive 
templates'' for the identification of the $\gamma$-ray sources. Geminga is the
best example of this source class in the EGRET catalog, and provides a template
of characteristics that includes behavior as a pulsar at X-ray and $\gamma$-ray
energies, but faint in optical wavelengths, with sporadic or no radio 
emission (see Caraveo, Bignami, \& Tr\"umper 1996 for a review). The 
question of whether there are other Geminga-like pulsars in the 3EG 
unidentified source catalog has often been raised. It has been suggested that
perhaps Geminga-like sources could account for the weaker mid-latitude 3EG
unidentified sources (Gehrels et al. 2000). On a case-by-case basis, the
multiwavelength strategy has been used to suggest isolated neutron star
counterparts to some EGRET sources. In fact, the identification of Geminga came
after a successful multiwavelength campaign, carried out over a 20 year
period (see Bignami \& Caraveo 1996 for a review). 
We describe a few other recent examples below. 

\subsection{The case of 3EG J1835+5918}

3EG 1835+5918 is the brightest and most accurately positioned unidentified 
EGRET source that has been persistently detected at high energy $\gamma$-rays 
(Nolan et al. 1996). 3EG J1835+5918 is located at high Galactic latitude at 
$l=88.74^\circ$, $b=25.07^\circ$, well away from the confusing diffuse 
emission. The source shows no strong evidence of variability 
(Reimer et al. 2000), and has a spectral index in the 70 MeV to 4 GeV range 
of $-1.7$ (Hartman et al. 1999). 
Despite its small error circle, 3EG J1835+5918 remained 
a mystery, and was the subject of several multifrequency studies 
(Reimer et al. 2000; Carrami\~nana et al. 2000; Mirabal et al. 2000;
Reimer et al. 2001; Mirabal \& Halpern 2001). No known flat-spectrum radio source 
was found in earlier searches of its error circle (Mattox et al. 1997). Its 
temporal and
spectral variability indicate that it is more similar to pulsars than blazars. 
We present the steps towards the identification of 3EG J1835+5918 in some
detail as a classic example of the use of the multifrequency strategy in the
identification of EGRET sources.  

\begin{figure} [t!]
\begin{center}
\includegraphics[height=18pc]{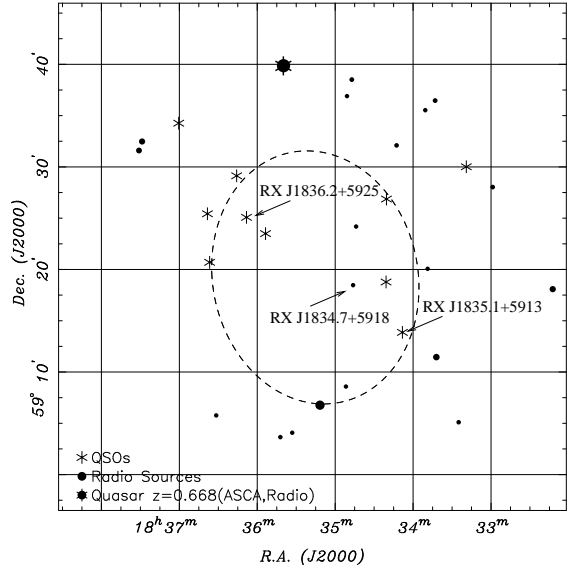}
\caption{Quasars and radio sources within the error circle (shown with dashed
lines) of 3EG J1835+5918. The positions of the three interesting objects,
described in the text, are individually marked (Mirabal et al. 2000). }
\end{center}
\label{1835sources}
\end{figure}

The error circle of 3EG J1835+5918 has been 
the subject of intense multiwavelength
study. Analysis of archival 
{\it ROSAT} HRI and PSPC as well as {\it ASCA} observations of the 
EGRET field yielded 
several point-like X-ray sources within the error circle of 
3EG J1835+5918 (Mirabal et al. 2000; Reimer et al. 2001). 
Optical identifications of the X-ray sources were carried out independently by 
Mirabal et al. (2000) and Carrami\~nana et al. (2000). 
Most of the sources were 
found to be either radio-quiet QSOs or coronal emitting stars or a galaxy
cluster. In addition, analysis of archival radio data (VLA, NRAO and WENSS) 
revealed 
only three sources within the 99\% error contour of 3EG J1835+5918, all of 
which were fainter than 4 mJy at 1.4 GHz. The positions of the quasars and 
radio sources in the vicinity of the EGRET source are shown in
figure~\ref{1835sources} (Mirabal et al. 2000). Three of the interesting
sources are individually marked in the figure: RX J1834.1+5913 is the brightest
quasar in the EGRET error ellipse, VLA J1834.7+5918 is the brightest of the
three weak radio sources within the EGRET error circle, and RX J1836.2+5925 is an
object that does not seem to have an optical counterpart. 
No blazar-like radio sources were found in the vicinity of the EGRET 
source. The brightest neighbouring radio sources were steep-spectrum radio 
galaxies or quasars. 

\begin{figure}[t!]
\begin{center}
\includegraphics[height=18pc]{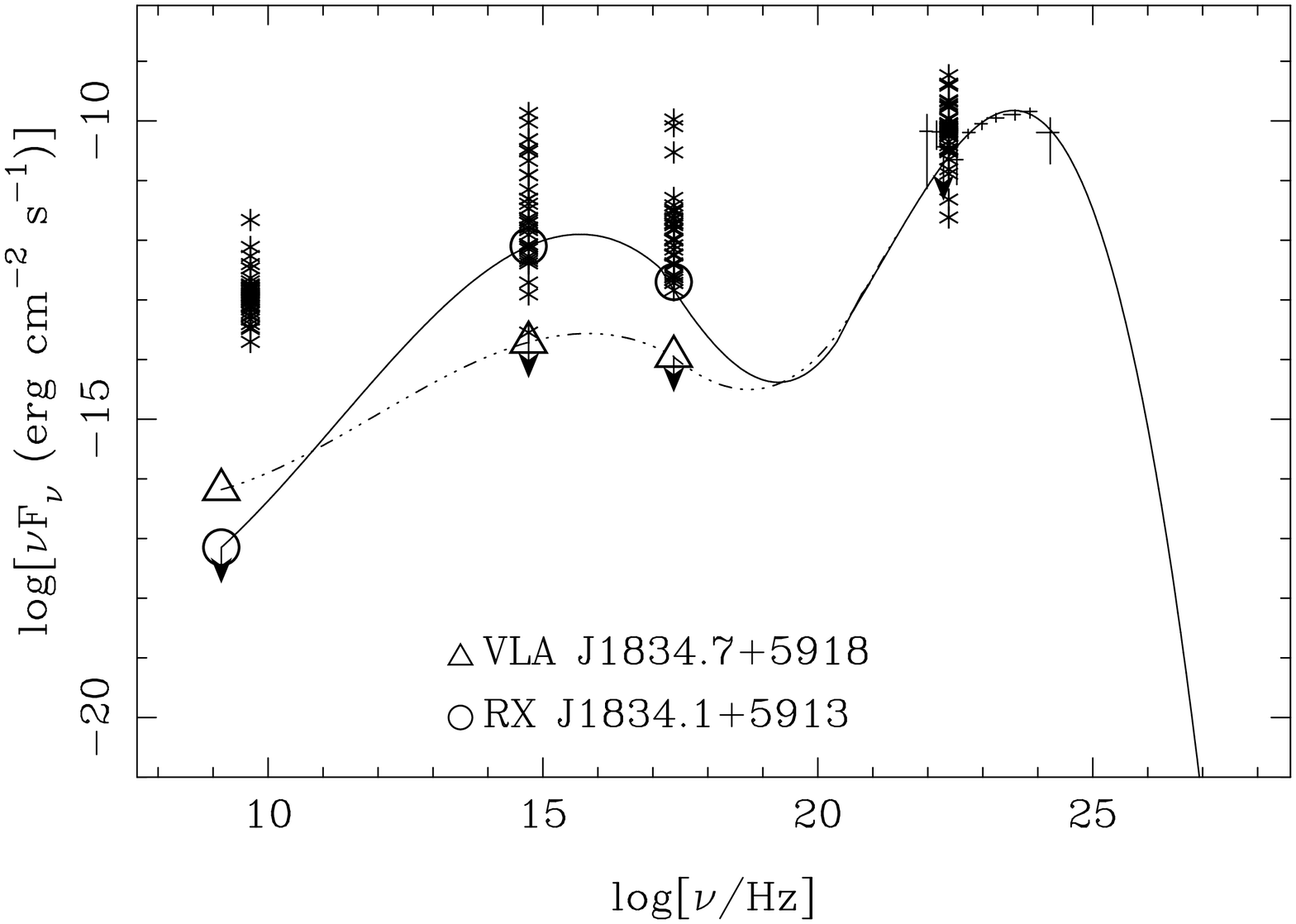}
\caption{The broadband fluxes of EGRET blazars, as compiled from the 
literature, are shown as asterisks. The X-ray and radio data of the two most
likely counterparts, RX J1834.1+5913 and VLA J1834.7+5918, if 3EG J1835+5918
were a blazar, are also shown. Note that the two candidates lie at the faint 
end of the distribution, making it unlikely that 3EG J1835+5918 is a blazar,
at least with properties similar to other EGRET-detected blazars 
(Mirabal et al. 2000). }
\end{center}
\label{1835sed}
\end{figure}

In fact, the broadband
characteristics of 3EG J1835+5918 were examined to see if they fall within the
multiwavelength parameters of the blazar class of sources seen by EGRET. Figure~\ref{1835sed} (Mirabal et al. 2000) shows the radio, optical, X-ray 
and $\gamma$-ray fluxes of the sample of well-identified blazars in Mattox et
al. (1997). For comparison, 
the fluxes of the brightest possible QSO counterpart, 
RX J1834.1+5913, and the most likely radio counterpart VLA J1834.7+5918 are 
also shown. A low energy synchrotron component and a high energy inverse 
Compton component is assumed.  Note that both 
the candidates are found to lie at the faint end of the distribution, making it
unlikely that 3EG J1835+5918 is a blazar. 

RX J1836.2+5925, indicated in figure~\ref{1835sources}, is the most intriguing
object within the error circle of 3EG J1835+5918. 
This object has no optical counterpart to a limit of 
$V> 25$ (Mirabal \& Halpern 2001), and has been suggested as a radio quiet pulsar,
and the most promising counterpart to the enigmatic $\gamma$-ray source  
3EG J1835+5918 (Reimer et al. 2001; Mirabal \& Halpern 2001). The ratio of the
$\gamma$-ray flux above 100 MeV of 3EG J1835+5918 to the X-ray flux (0.12 - 2.4
keV) of RX J1836.2+5925 is similar to that of other similar candidates
considered to be of pulsar origin (Reimer et al. 2001). The lack of an optical
counterpart, and the non-variability of the $\gamma$-ray source are all
characteristic signatures for a radio-quiet pulsar. 

Recently, Mirabal \& Halpern 
(2001) have presented arguments that RX J1836.2+5925 is 
indeed a neutron star, and could be a nearby, rotation-powered radio-quiet
$\gamma$-ray 
pulsar.  Although its X-ray flux is at least 10 times fainter than that of
Geminga, RX J1836.2+5925 is possibly older or more distant than Geminga, and 
the most likely counterpart of 3EG J1835+5918. 

\begin{figure}[t!]
\begin{center}
\includegraphics[height=18pc]{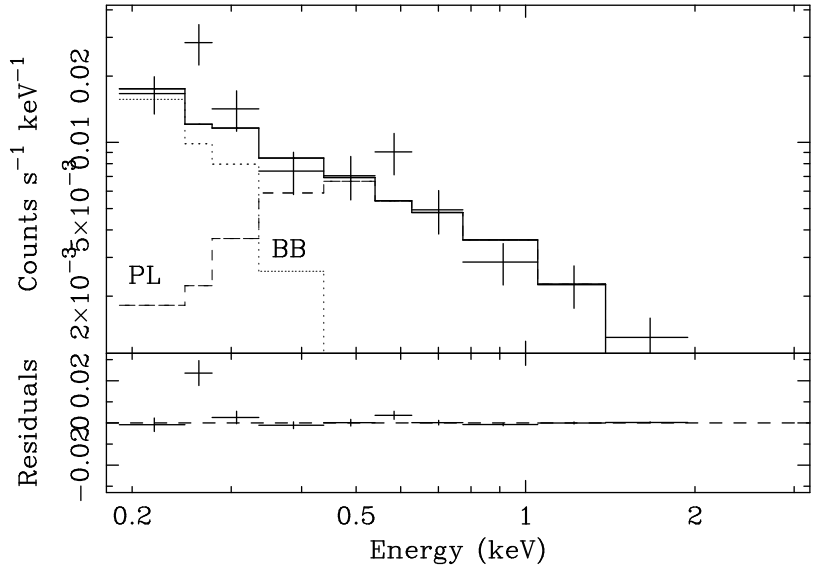}
\caption{{\sl Chandra} ACIS-S3 spectrum of RX J1836.2+5925, the neutron star
counterpart of 3EG J1835+5918. Data shown as crosses; best fit model as a thick
line, with contributions of a blackbody (BB) and power-law (PL) components. 
Difference between data and model is shown in the bottom panel. Figure from 
(Halpern et al. 2003). }
\end{center}
\label{chandra1835}
\end{figure}

Using deep {\sl Chandra} data, along with HST and radio observations of RX J1836.2+5925, Halpern et
al. (2003) have presented further, conclusive evidence that an older, possibly
more distant Geminga-like pulsar is responsible for the origin of $\gamma$-rays
from 3EG J1835+5918. Figure~\ref{chandra1835} shows the {\sl Chandra} ACIS-S 
spectrum of RX J1836.2+5925 with a fit that requires a two-component model: a
thermal blackbody of $T_\infty\simeq3\times 10^5$ K with a power law component 
of photon index $\Gamma\simeq2$. This non-thermal extension to the X-ray
spectrum is characteristic of the EGRET pulsars and further supports the
identification of the EGRET source. 

\subsection{Other neutron star candidates}

Two other examples of neutron star candidates are 3EG J0010+7309 (Brazier et
al. 1998) and 2EG J2020+4026 (3EG J2020+4026) (Brazier et al. 1996). 
3EG J2020+4026 is coincident with the $\gamma$-Cygni supernova remnant
G78.2+2.1. Brazier et al. (1996) studied {\sl ROSAT} PSPC data on this region,
and found a single, point-like X-ray source in the EGRET 95\% error contour, 
RX J2020.2+4026, the only plausible counterpart. The flux ratio at 
$\gamma$-ray and X-ray energies, $F_\gamma/F_X$ was found to be $\sim 6000$,
similar to that of Geminga, and very different from non-pulsar sources. Brazier
et al. suggested RX J2020.2+4026 as the counterpart of the EGRET source, and 
3EG J2020+4026 as possibly a young pulsar.  No radio source was found at the
position of the X-ray source, and it is likely that the pulsar is Geminga-like. 

3EG J0010+7309 (2EG J0008+7307) is a similar example of a possible
pulsar/neutron star candidate. 
It has a smaller error box compared to most EGRET
unidentified sources because it was clearly visible above 1 GeV. 
The source is spatially coincident  with the supernova
remnant CTA~1. X-ray data ({\sl ROSAT} and {\sl ASCA}) of this region was studied by Seward et
al. (1995) and Slane et al. (1997) and re-examined by Brazier et al. (1998) in some detail in an
effort to identify 3EG J0010+7309. The small EGRET error contour encloses one
{\sl ROSAT} source, RX J0007.0+7302, which Brazier et al. (1998) suggest as the
counterpart of the EGRET source. The X-ray source is believed to be a pulsar. 
No optical counterpart was found for the X-ray source, suggesting a neutron 
star nature of the source. No radio pulsed source was detected at the position
of the X-ray source either. Brazier et al. (1998) suggest that the X-ray source
is possibly a radio quiet pulsar. 

\section{Young pulsar candidates}

The majority of the identified EGRET sources at low Galactic latitudes are
pulsars. In view of this fact, pulsars are a natural template for counterpart
searches for the Galactic plane 3EG sources (see Caraveo 2002 for a review). 
In fact, there have been several
efforts for pulsar searches at radio wavelengths (e.g. Nice \& Sayer 1997; 
Nel et al. 1996). Other studies using Parkes data have yielded radio pulsar
candidates for 3EG J1420-6038, 3EG J1837-0606 (D'Amico et al. 2001) and 3EG
J1013-5915 (Camillo et al. 2001). 
Torres, Butt \& Camilo (2002) have recently reported on a correlative study
between the low latitude 3EG sources and the newly discovered pulsars in the
Parkes multibeam radio survey (Manchester et al. 2001), confirming earlier
studies, but not yielding any new counterparts. Other possible associations
include PSR B1046-58 with 3EG J1048-5840 (Kaspi et al. 2000; Thompson 2001).  
Confirmation of $\gamma$-ray pulsars will only come from measuring the timing
charateristics at $\gamma$-ray energies, and will be a priority for future
$\gamma$-ray missions such as GLAST. Here we discuss a couple of individual cases
where extensive multifrequency efforts have been utilized to suggest pulsar
counterparts for 3EG sources. 

\subsection{3EG J2021+3716: The young radio pulsar PSR J2021+3651}

\begin{figure}[t!]
\begin{tabular}{cc}
\begin{minipage}{0.46\linewidth}
  \begin{center}
     \includegraphics[height=15pc]{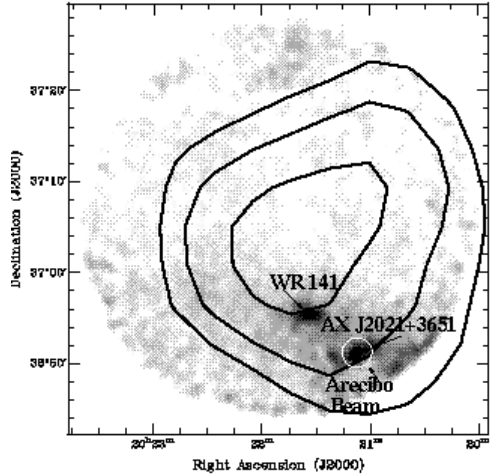}
  \end{center}
\vspace{-0.5cm}
  \caption{$ASCA$ GIS image (2-10 keV) of the error box of the $\gamma$-ray
source GeV J2020+3658. The contours correspond to 68\%, 95\% and 99\% confidence
levels. The position of the $ASCA$ unidentified hard X-ray source, suggested as
the counterpart of the EGRET source, is shown. The circle corresponds to the
$3^\prime$ Arecibo beam. Figure from Roberts et al. (2002). }
\label{2021asca}
\end{minipage}&
\begin{minipage}{0.46\linewidth}
  \begin{center}
    \includegraphics[height=9pc]{mukherjee_f13.eps}
  \end{center}
  \caption{Pulsar profile of PSR J2021+3651 at 1.4 GHz. Figure from Roberts et
al. (2002)}
\label{2021pulsar}
\end{minipage}
\end{tabular}
\end{figure}

This is a classic example of how multiwavelength studies of the EGRET field of
3EG J2021+3716 (GeV J2020+3658) 
was used to suggest a pulsar counterpart for the EGRET source. Along
with 3EG J2016+3657, this source is possibly associated with the COS-B source
2CG 075+00. Both sources were discussed earlier 
in \S 2.1, where we described multifrequency
studies leading to the identification of 3EG
J2016+3657 with a blazar counterpart. Figure~\ref{2016rosatpspc} shows the {\sl ROSAT} PSPC data covering
the error boxes of the two EGRET sources, as well as the contour of the COS-B
source. Roberts et al. (2002) have recently reported on multiwavelength studies
of GeV J2020+3658 in which they carried out a deep search 
for radio pulsations toward the
unidentified {\sl ASCA} source AX J2021.1+3651 in the error box of the EGRET
source. AX J2021.1+3651 is one of the hard X-ray sources listed in 
the
{\sl ASCA} catalog of potential X-ray counterparts of GeV sources, a catalog
resulting from X-ray studies of the EGRET fields (Roberts, Romani, \& Kawai
2001). Figure~\ref{2021asca} shows the {\sl ASCA} GIS image of the
$\gamma$-ray source region. Roberts et al. (2002) observed AX J2021.1+3651 with the Wideband Arecibo Pulsar
Processor (WAPP) and discovered a new young and energetic pulsar PSR
J2021.1+3651, which they argue is the counterpart to the EGRET source GeV
J2020+3658. WR 141 is a Wolf-Rayet star also in the field of view. Figure~\ref{2021pulsar} shows the 1.4 GHz pulse profile of PSR
J2021+3651. The positional coincidence of the pulsar with GeV J2020+3658, the
hard spectrum of the EGRET source, and its low variability, and the fact that
Roberts et al. (2002) find high inferred spin-down luminosity for the pulsar
strongly argue that the two sources are related. Confirmation of the
identification will hopefully  come in the future with GLAST observations. 

\subsection{The case of 3EG J2227+6122}

3EG J2227+6122 is another source at low Galactic latitude ($l=106.^\circ5$,
$b=3.^\circ2$) that was the subject of recent multiwavelength study 
(Halpern et al. 2001b). X-ray, radio, and optical
observations of the 
EGRET field together point to the possibility that 3EG J2227+6122 is most 
likely a young, energetic pulsar, with an associated  X-ray pulsar wind nebula (PWN), 
enclosed in a small non-thermal radio shell. 

\begin{figure}[t!]
\begin{minipage}{0.96\linewidth}
\begin{center}
\includegraphics[height=15pc]{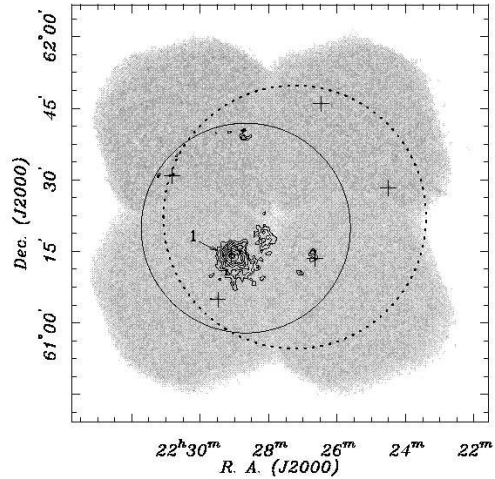}
\end{center}
\caption{Composite ROSAT HRI image of the 3EG~J2227+6122 field. The 
dashed circle corresponds to the 95\% error contour of the EGRET source. 
Except for \# 1, all the X-ray point sources (plus signs) are 
bright stars. \# 1 is the only unidentified HRI source, and is coincident with 
a bright, hard source seen in the ASCA GIS image (contours). The solid circle 
corresponds to the ASCA GIS field. Figure from Halpern et al. (2001b).}
\label{rosathri2227}
\end{minipage}
\end{figure}

Figure~\ref{rosathri2227} shows a composite {\it ROSAT} HRI image of the error circle
of 3EG J2227+6122, showing 6 point-like X-ray sources within the EGRET 95\%
contour (Halpern et al. 2001b). All sources, except \# 1 have optical
spectroscopic identifications obtained 
using the KPNO 2.1~m telescope and Goldcam
spectrograph, and are either bright K and M type stars, or emission-line 
stars. Source \# 1, RX J2229.0+6114, also 
detected in the {\it ASCA} image of the region was found to have no optical 
counterpart. The contours in figure~\ref{rosathri2227}
correspond to the {\it ASCA} GIS image of the source AX J2229.0+6114. The X-ray source 
RX/AX J2229.0+6114 was found to have a non-thermal spectrum with a power law photon index 
$\Gamma=1.51\pm 0.14$.  

\begin{figure}
\begin{minipage}{0.96\linewidth}
\begin{center}
\includegraphics[height=15pc]{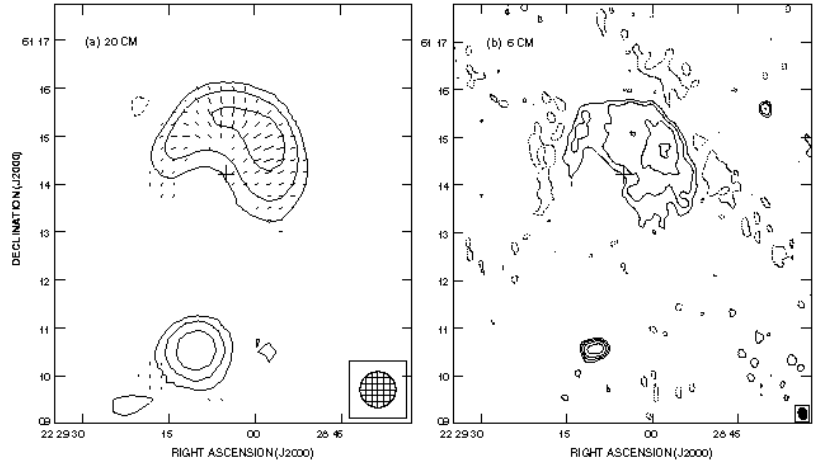}
\caption{(Right) 20 cm NVSS map and (left) 6 cm VLA map, showing the
shell-like radio source. The position of the X-ray RX J2229.0+6114 source is
shown with a  `+'. Notice the polarization vectors in the 20 cm NVSS
map. Figure from Halpern et al. (2001b).}
\end{center}
\label{radio2227}
\end{minipage}
\end{figure}

Halpern et al. (2001b) obtained 20 cm (NVSS) and 6 cm (VLA) images of the 
error circle of the EGRET source. These images are shown in
figure~\ref{radio2227}. Interestingly, they found that there was 
only one radio source that was coincident with an X-ray source in the field, 
and that was source \# 1, RX J2229.0+6114 of figure~\ref{rosathri2227}. 
The radio source VLA J2229.0+6114 has an 
incomplete circular shell-like structure, with a high degree of linear
polarization evident throughout the shell. Halpern et al. (2001b) have 
presented convincing arguments suggesting that VLA J2229.0+6114 and 
RX/AX J2229.0+6114 are associated with each other. 

\begin{figure}
\begin{tabular}{cc}
\begin{minipage}{0.46\linewidth}
  \begin{center}
     \includegraphics[height=10pc]{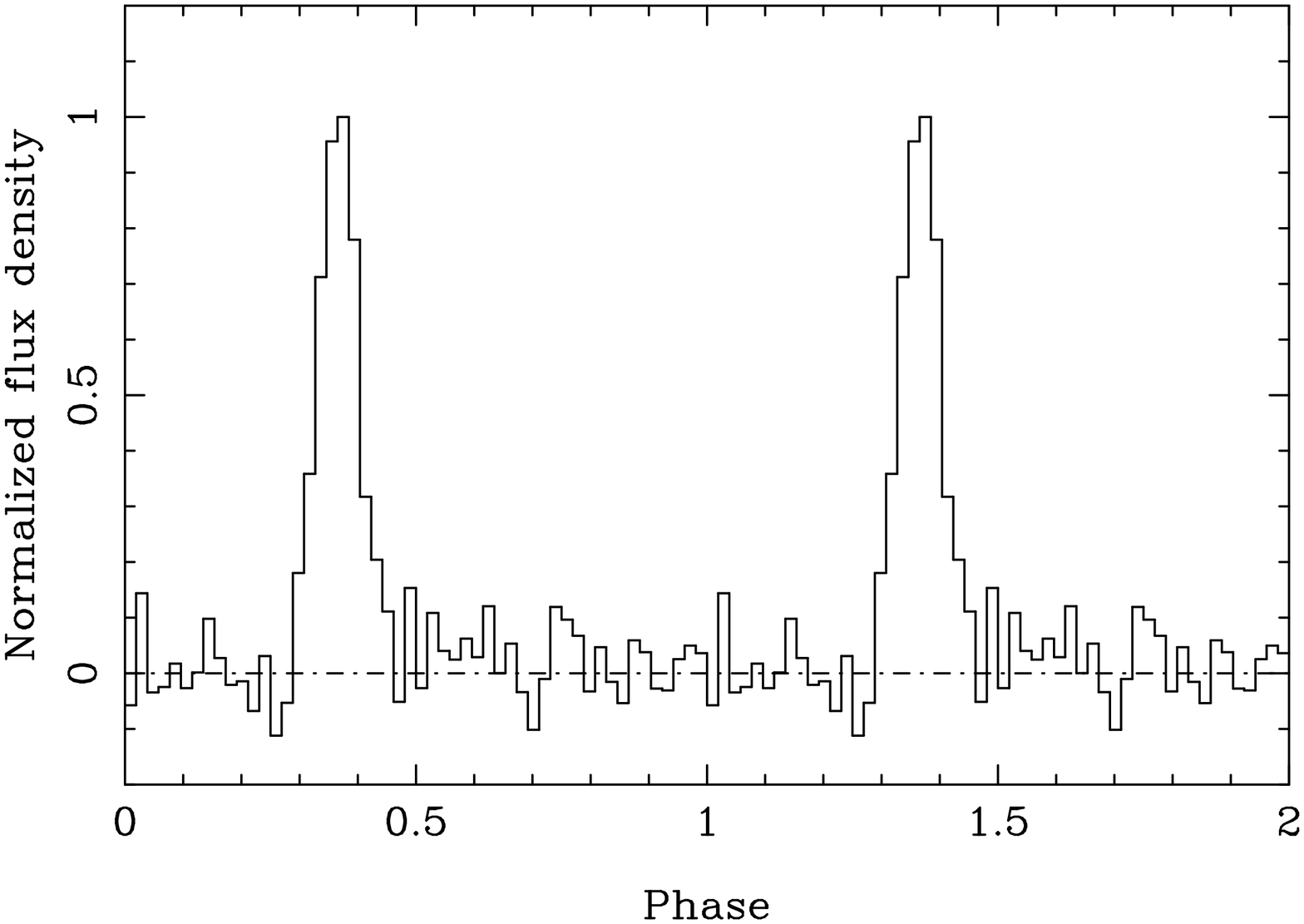}
  \end{center}
  \caption{Radio pulse profile of PSR J2229+6114 at 1412 MHz observed with the
Lovell radio telescope at Jodrell Bank. Figure from Halpern et al. (2001c). }
\label{2229psr_radio}
\end{minipage}&
\begin{minipage}{0.46\linewidth}
  \begin{center}
\vspace{-0.5cm}
    \includegraphics[height=10pc]{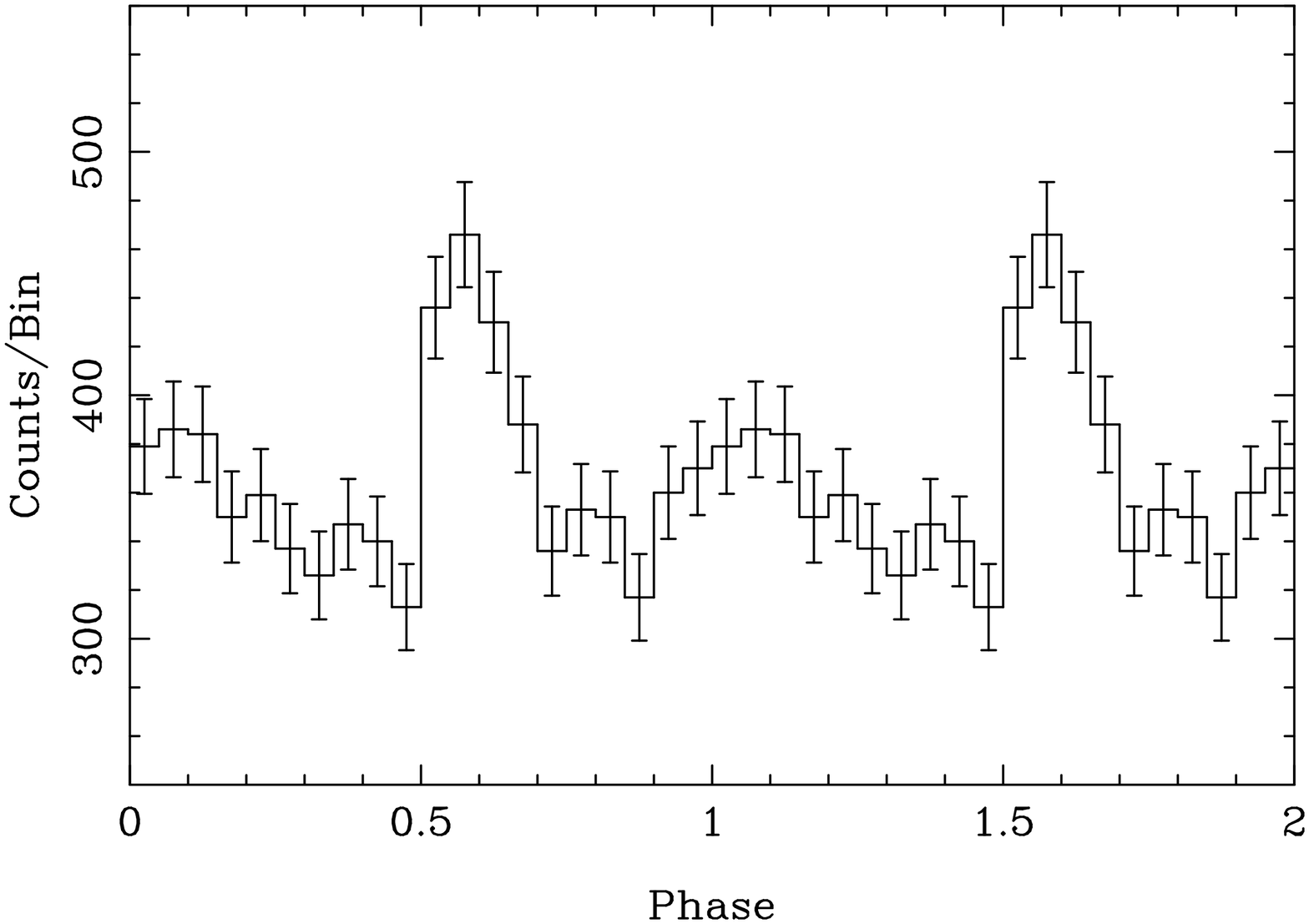}
  \end{center}
  \caption{X-ray pulse profile of PSR J2229+6114 in the 0.8-10 keV band from
  {\sl ASCA} GIS. Figure from Halpern et al. (2001c).}
\label{2229psr_xray}
\end{minipage}
\end{tabular}
\end{figure}

Recently, Halpern et al. (2001c) described further multiwavelength 
observations of the X-ray source RX/AX J2229.0+6114
with the {\sl Chandra} imaging CCD array ACIS-I, and at radio frequencies, and
reported on the detection of radio and X-ray pulsations at a period of 51.6 ms 
from the X-ray
source. The {\sl Chandra} image clearly shows a point
source surrounded by diffuse emission. Halpern et al. note  that this
morphology, together with the 
non-thermal spectrum of the X-ray nebula 
indicates a ``composite'' supernova 
remnant, which they have called G106.6+2.9. Figure~\ref{2229psr_radio}
shows the radio pulse profile of PSR J2229+6114 at 1412 MHz, observed with the
Lovell radio telescope at Jodrell Bank in 2001 February. Following the radio
pulsar discovery, Halpern et al. (2001c) searched the {\sl ASCA} GIS data for
X-ray pulsations. Their results, shown in figure~\ref{2229psr_xray}, indicate a
pulsed fraction of 22\%. 

These observations leave very little doubt that the EGRET source 3EG J2227+6122
is indeed the young and energetic 51.6 ms X-ray/radio pulsar PSR J2229+6114. 
Further confirmation will be possible after the launch of GLAST and if direct
pulsations are observed at $\gamma$-ray energies. 

\section{Other source classes}

Multiwavelength studies have led to the tentative identification of EGRET
sources with counterparts from some other source classes. For example, 
EGRET sources have long been associated with supernova remnants (SNRs), 
and there are
several examples of positional coincidences between EGRET sources and SNRs (see
Torres et al. 2003b for a review). Examples are IC 443 (GeV J0617+2237), W 28
(GeV J1800-2328), among others. 
 
Similarly, there have been at least a couple of examples where a particular 
EGRET source has been associated with a microquasar. Paredes et al. (2000) have
suggested that the microquasar LS 5039 is the counterpart to the EGRET source
3EG J1824-1514. The unidentified EGRET source 3EG J1828+0142 has been suggested
as a possible Galactic microblazar, 
following multifrequency studies at X-ray and radio energies (Butt et
al. 2002). Several authors have suggested that some variable $\gamma$-ray
unidentified sources in the Galactic plane could be interpreted as 
microquasars 
(e.g. Paredes et al. 2000; Romero 2001; 
Kaufman-Bernado, Romero, \& Mirabel 2002). 
It is possible that microblazars are a ``new'' class of $\gamma$-ray
sources - further confirmation will come after more such sources are identified
at $\gamma$-ray energies in the future. 

In a few cases EGRET sources have been identified with peculiar binary systems,
following a multiwavelength study of the EGRET error box. 
Two possible examples are 3EG J0634+0521 (2CG 135+01) associated with the 
binary system 
of a compact object and a Be star companion, 
SAX J0635+0533 (Kaaret et al. 1999), and 3EG J0241+6103, associated with the
periodically variable radio/Be/X-ray/ source GT 0236+610/LSI $+61^\circ 303$ 
(Bignami et al. 1981). Another example is the possible association of 3EG
J0542+2610 with the Be/X-ray transient A0535+26 (Romero et al. 2001). 
The gamma-ray production
mechanism proposed in this case, based on the
magnetosphere model of Cheng \& Ruderman (1989), could explain emission from 
other variable EGRET sources in the plane. 

The possible association of the variable radio star LSI $+61^\circ 303$ with the COS-B
$\gamma$-ray source 2CG 135+01, was first noted by Bignami et al. (1981), who
described the Einstein X-ray identification of the source. 
EGRET observations of 2CG 135+01 (3EG J0241+6103), a prominent unidentified
source near the Galactic plane, were presented by Tavani 
et al. (1998). LSI $+61^\circ 303$ was the subject of a multiwavelength investigation at
radio, optical, infrared and hard X-ray/$\gamma$-ray frequencies, in order to
confirm the association of the with the $\gamma$-ray source (Strickman et
al. 1998). 
Although there was no conclusive proof of the identification, this is an
intriguing association. 

\begin{figure}[b!]
  \begin{center}
     \includegraphics[height=10pc]{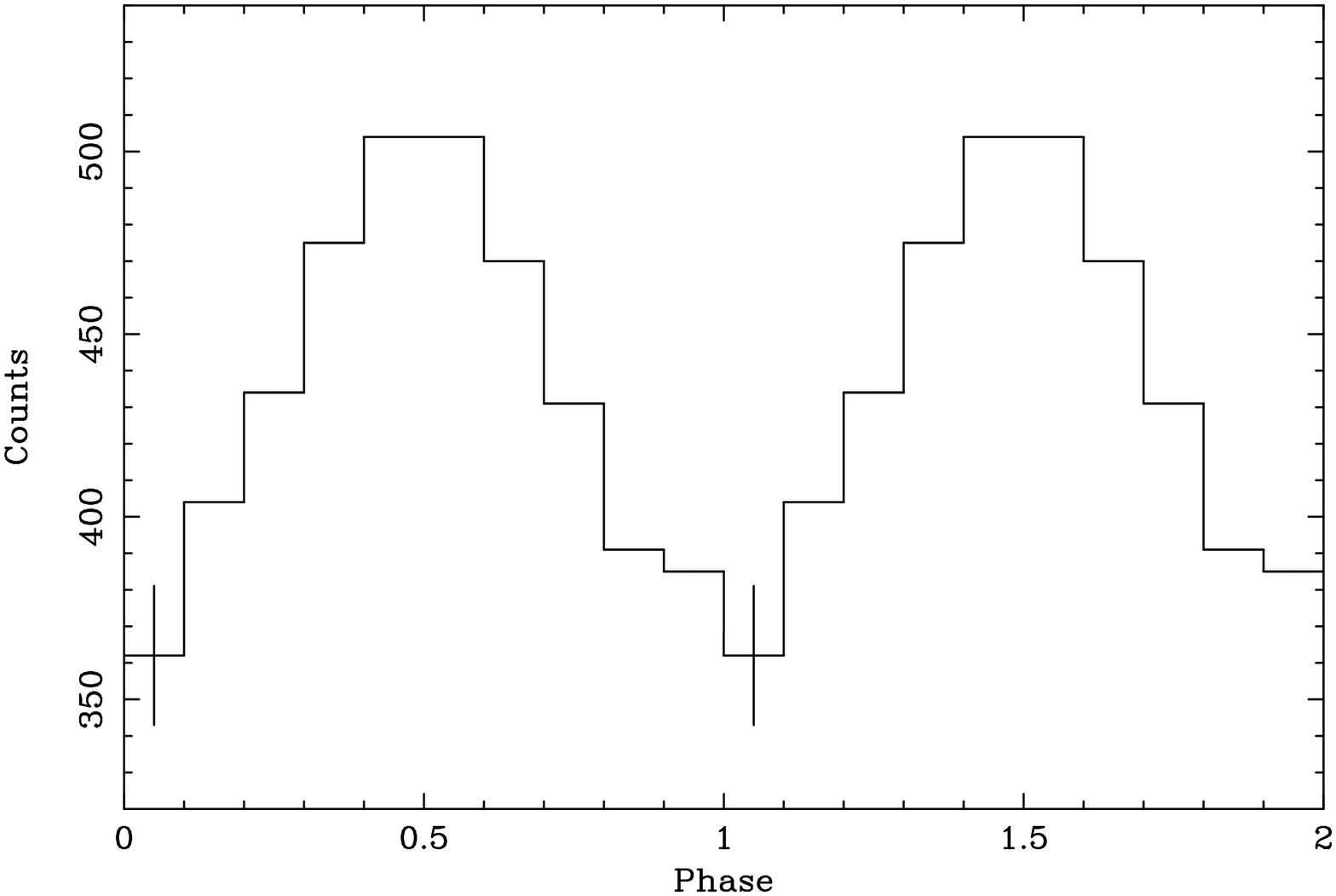}
  \end{center}
  \caption{X-ray pulse profile of SAX J0635+0533, associated with the EGRET
     source 3EG J0634+0521. The X-ray source is believed to be X-ray binary,
     emitting $\gamma$-rays. Figure from Cusumano et al. (2000). }
\label{0635pulse}
\end{figure}

The second such example is the case
of 3EG J0634+0521, associated with the hard spectrum X-ray source, SAX
J0635+0533, 
discovered in the error box of the EGRET source (Kaaret et al. 1999). 
Optical observations of SAX J0635+0533 in the $V$ band showed a counterpart with
broad emission lines, and the colors of an early B type star. Subsequent
discovery of pulsations at a period of 33.8 ms from the SAX source further
strengthens the association (Cusumano et al. 2000). Figure~\ref{0635pulse} shows
the pulse profile obtained by analyzing the SAX data. 
The pulsations were suggested to be due to a neutron star in a binary system
with a Be companion. A definitive proof of the association with the EGRET source
will require detection of periodicity in the EGRET source. 

\begin{figure}[t!]
  \begin{center}
     \includegraphics[height=20pc]{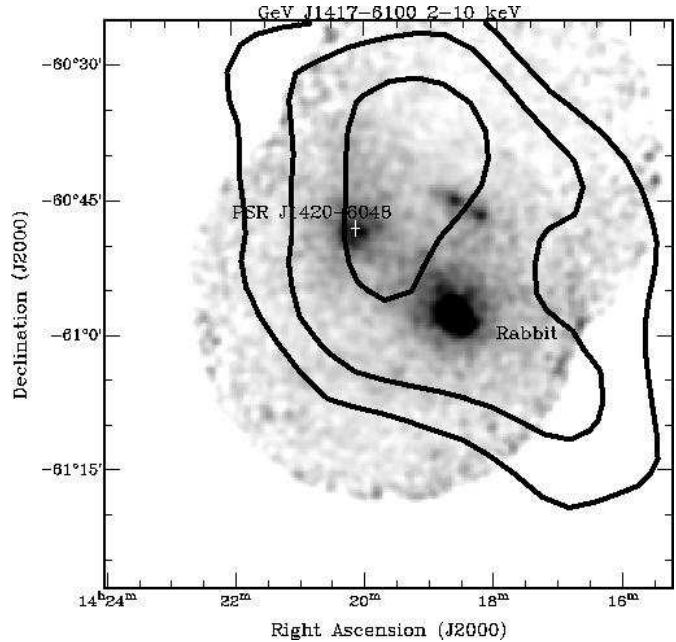}
  \end{center}
  \caption{ $ASCA$ image of GeV J1417-6100 showing the two extended hard X-ray
     sources. Figure from Roberts et al. (2001). }
\label{1417mallory_xray}
\end{figure}

\begin{figure}[t!]
  \begin{center}
     \includegraphics[height=20pc]{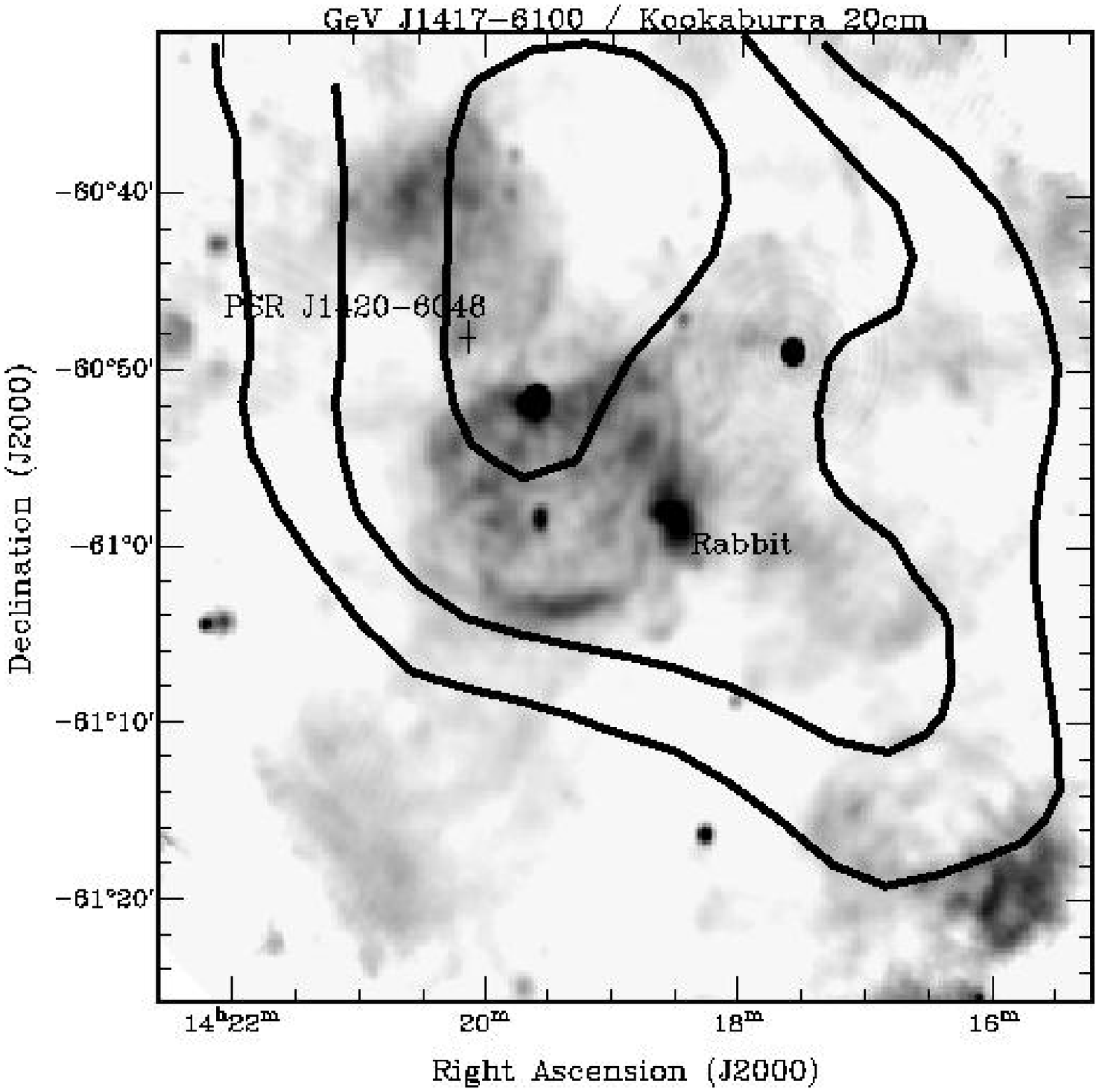}
  \end{center}
  \caption{ 20 cm ATCA image of GeV J1417-6100, showing the Kookaburra
     Nebula. The locations of the  pulsar PSR
     J1420-6049 and the Rabbit Nebula are indicated. Figure from Roberts et al. (2001). }
\label{1417mallory_radio}
\end{figure}

Several studies have suggested that some variable 3EG sources are 
associated with pulsar wind nebulae (see Roberts, Gaensler, \& Romani 2002 for a
review). X-ray and radio studies of pulsar wind nebulae (PWN) suggest that
several of these sources are associated with unidentified EGRET sources. 
One example is that of GeV J1417-6100 (3EG J1420-6038), the error box of which
coincides with the Kookaburra radio complex, within which are two extended
hard X-ray sources (Roberts et al. 1999). 
Figure~\ref{1417mallory_xray} shows the image of 
GeV J1417-6100
in X-rays (Roberts et al. 2001), with the two X-ray sources
indicated. Figure~\ref{1417mallory_radio} shows the radio 20 cm image of the
region, showing the Kookaburra Nebula and the location of the Rabbit Nebula at
the edge of the Kookaburra complex. One of the two hard X-ray sources in the
field (AX J1420.1-6049) 
was recently found to contain the 68 ms radio 
pulsar PSR J1420-6048 (D'Amico et al. 2001),
suggesting that the source is an X-ray and radio PWN.  Recently Roberts, Romani
\& Johnston (2001) have
presented multiwavelength X-ray, radio, and infrared observations of the pulsar
and the surrounding nebula. PSR J1240-6048 is a possible counterpart of the
$\gamma$-ray source. 
Roberts, Gaensler \& Romani (2002) name a few other variable 
EGRET unidentified sources associated with PWN, GeV J1825-1310, GeV J1809-2327,
suggesting that some variable EGRET 3EG sources may be PWN. Another possible
example is that of 3EG J1410-6147. Doherty et al. (2003) have presented radio
continuum, HI and X-ray ({\sl Chandra}) observations of this field recently. The
EGRET source could be a PWN, near the pulsar PSR J1412-6145, but the association
is not definite. 

\section{Studies of EGRET unidentified sources at TeV energies}

A new window for the observations of unidentified EGRET sources is now 
available at energies above 300 GeV by using 
imaging atmospheric Cherenkov (IACT) detectors, 
and at lower energies ($\sim 100$ GeV), 
solar array experiments like STACEE and CELESTE. 
IACTs have the advantage of superior sensitivity and angular
resolution. IACTs have successfully detected high energy $\gamma$-ray
emission from both Galactic \& extragalactic objects at energies above 300 GeV
(see Ong 2003 for a review of the experiments and recent 
results).  The Whipple 10 m telescope has been used to observe several
unidentified EGRET sources in the past (Buckley et al. 1997), but none was
detected. Recently, Fegan (2001) reported on the upper limits  of a selected
number of EGRET unidentified sources observed with Whipple. 

\begin{figure}[t!]
  \begin{center}
     \includegraphics[height=20pc]{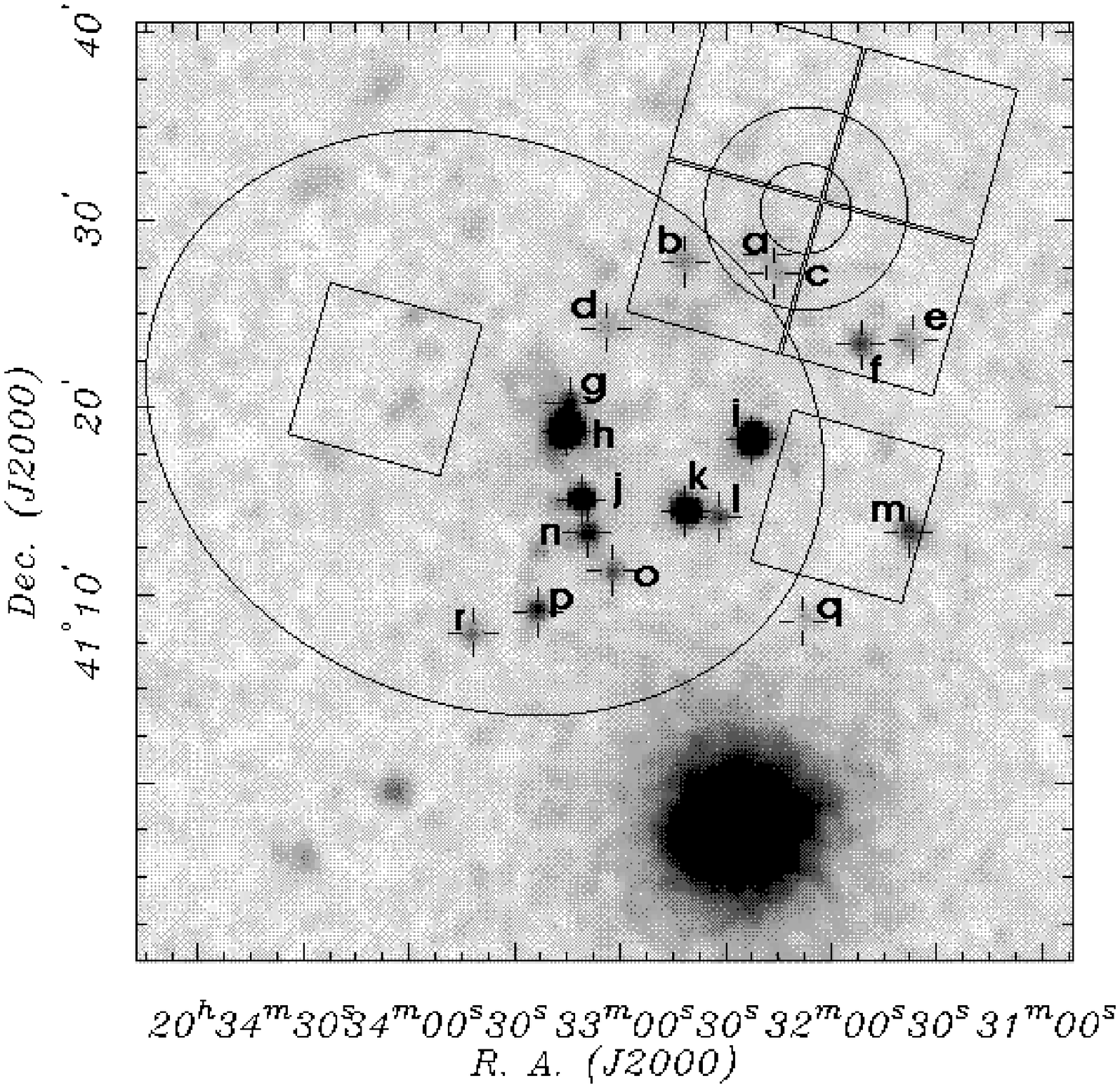}
  \end{center}
  \caption{$ROSAT$ PSPC X-ray image of the error circles of 3EG J2033+4118/TeV
     J2032+4130. The ellipse is the 95\% uncertainty location of 3EG J2033+4118
from Mattox et al. (2001).  The small circle is the $1\sigma$ uncertainty of
the centroid of TeV J2032+4130, and the large circle
is the estimated Gaussian $1\sigma$ extent of the TeV emission (Aharonian et al. 2002).
The bright source is Cyg X-3.
The properties of the marked sources are described in Mukherjee et al. (2003).
The squares are the fields of view of the CCDs in the subsequent
$Chandra$ observation (see Figure~4). Figure from Mukherjee et al. (2003). }
\label{2033rosat}
\end{figure}

A recent exciting news was the detection of what might be the first 
unidentified ``TeV'' source in the Cygnus region.  This source 
was
detected serendipitously by the HEGRA CT-System (Aharonian et al. 2002) in
observations originally devoted to the EGRET unidentified source 
3EG J2033+4118
and Cygnus X-3 and is known as TeV J2031+4130. The error circle of TeV J2033+4130 overlaps that of the EGRET
unidentified source, but the two are not necessarily related. 
Figure~\ref{2033rosat} shows an X-ray image taken with {\it ROSAT\/}
PSPC in the energy range 0.2--2.0 keV, covering the field of
3EG J2033+4118/TeV J2032+4130. This TeV/EGRET field was the subject of recent 
multiwavelength study, with the intent of searching for a counterpart for the
TeV source (Mukherjee et al. 2003; Butt et al. 2003). Most of the brighter X-ray point
sources in Figure~\ref{2033rosat} were observed optically and identified 
spectroscopically to be  a mix of early and late-type stars, unlikely to be
counterparts of the $\gamma$-ray source (Mukherjee et al. 2003). 

\begin{figure}[t]
  \begin{center}
    \includegraphics[height=20pc]{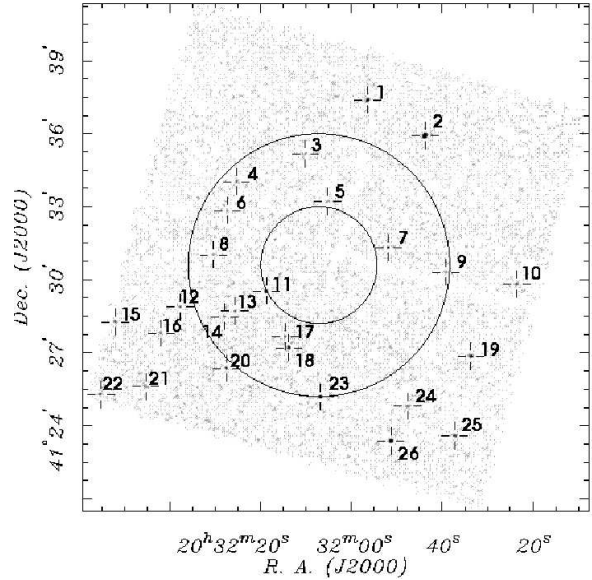}
  \end{center}
\vspace{-0.5cm}
  \caption{$Chandra$ ACIS-I image of the field of TeV J2032+4130.
The properties of the
numbered sources are given in Mukherjee et al. (2003). The small circle is the $1\sigma$ uncertainty of
the centroid of TeV J2032+4130, and the large circle
is the estimated Gaussian $1\sigma$ extent of the TeV emission (Aharonian et al. 2002).
The brightest $Chandra$ source \#2 was not detected in $ROSAT$ 
images. Figure from Mukherjee et al. (2003). }
\label{2033chandra}
\end{figure}

Recently, in August 2002 {\it Chandra} made a 5 ks
director's discretionary observation (Butt et al. 2003) 
of the field of TeV J2032+4130. Figure~\ref{2033chandra} shows the 
{\it Chandra} image of TeV J2032+4130, with the brightest point
sources marked (Mukherjee et al. 2003). Optical imaging observations of 
these sources show that they are mostly stars, or in some cases 
non detected, probably 
AGNs that are highly absorbed by the Galactic ISM. Mukherjee et al. (2003) 
draw attention to source \# 2 in figure~\ref{2033chandra}, the brightest X-ray
source in the {\sl Chandra} image, and a transient that is missing from 
the {\sl
ROSAT} image of the region. The hard X-ray spectrum, rapid variability, and red optical/IR colors of this object
suggest that it is a distant, quiescent X-ray binary system. It is too early to
tell whether this X-ray source is associated with the $\gamma$-ray source. 
It is possible that TeV J2032+4130 is an extended source. In this case the TeV emission
would not necessarily be centered on any point source counterpart at other 
wavelengths. Aharonian et al. (2002) have hypothesized
two possible origins for extended TeV emission. 
One is that TeV emission could arise from $\pi^0$ decay resulting from
hadrons accelerated in shocked OB star winds and interacting with a local,
dense gas cloud.  The other is inverse Compton TeV emission
in a jet-driven termination shock, either from an as-yet undetected 
microquasar, or from Cyg X-3. Butt et al. (2003) have also presented a {\sl
Chandra}/VLA follow-up of TeV J2032+4130, and have argued that the TeV 
source is an extended one that is not detected yet in radio or X-ray. 
Future observations with {\sl Chandra} and GLAST as well as by IACTs will help 
resolve the nature of this source. 

\subsection{Summary and future directions}

One of the most intriguing questions raised by EGRET has been the nature of its
unidentified sources. 
Resolving the mystery of the unidentified EGRET sources remains a daunting 
task, even after much study. We have presented a selective 
review of some recent work done towards understanding the
nature of the unidentified EGRET sources using a multiwavelength approach. 
This strategy of identification, although a systematic method, is a time 
consuming process requiring detailed
multifrequency studies of EGRET fields. It is, however, a promising method, 
and has yielded several new source identifications in the past
decade. The identification process is hampered by the large error boxes of 
EGRET sources. In the future, with smaller $\gamma$-ray error circles promised
by GLAST, this strategy should secure more confident source identifications. 
Future observations with GLAST or AGILE will also enable us to determine 
$\gamma$-ray 
source positions more accurately, and perhaps search for pulsations directly in
the $\gamma$-ray data. 

Although progress has been made in the identification of individual 
EGRET sources, both by looking at the sources as a group and by doing follow-up
multifrequency observations on a case-by-case basis, 
the majority of the EGRET (3EG) sources remain unidentified. 
It is possible that 
there may be a new class of $\gamma$-ray emitters, yet to be identified, made
up of several of the EGRET sources. 
In the interim before GLAST or AGILE, several of the
unidentified EGRET sources will continue to be observed above 250 GeV by 
ground-based instruments like VERITAS, as well as by 
the new generation low-threshold ground-based Cherenkov detectors like 
STACEE (Covault et al. 2003) and CELESTE (Nuss et al. 2003), sensitive 
to energies as low as 50 GeV. In the 
future, unidentified 3EG sources are likely to be studied not only by 
satellite-based experiments like GLAST, but also by next generation
ground-based detectors like VERITAS (Ong et al. 2003), MAGIC (Martinez et
al. 2003), HESS (Hofmann et al. 2003) and CANGAROO (Ohishi et al. 2003). 
Detection of very high energy
$\gamma$-ray emission from unidentified 3EG sources with ground-based 
atmospheric Cherenkov telescopes is likely to open an exciting new chapter in
the study of these sources.  
\bigskip

This research was supported in part
by the National Science Foundation.

\end{document}